\documentclass[twocolumn,english,superscriptaddress,floatfix,longbibliography]{revtex4-1}
\usepackage[T1]{fontenc}
\usepackage[utf8]{inputenc}
\usepackage{textcomp}
\usepackage{amsmath}
\usepackage{amssymb}
\usepackage{graphicx}

\makeatletter

\providecommand{\tabularnewline}{\\}

\usepackage{upgreek}
\usepackage{braket}

\makeatother

\usepackage{babel}
\begin{document}
\title{Light-controlled THz plasmonic time-varying media: momentum gaps,
entangled plasmon pairs, and pulse induced time-reversal}
\author{Egor I. Kiselev}
\affiliation{Department of Physics, Technion, Haifa 3200003, Israel}
\affiliation{The Helen Diller Quantum Center, Technion, Haifa 3200003, Israel }
\author{Yiming Pan}
\affiliation{School of Physical Science and Technology and Center for Transformative
Science, ShanghaiTech University, Shanghai 200031, China}
\author{Netanel H. Lindner}
\affiliation{Department of Physics, Technion, Haifa 3200003, Israel}
\affiliation{The Helen Diller Quantum Center, Technion, Haifa 3200003, Israel }
\begin{abstract}
This letter establishes a Floquet engineering framework in which coherent
high frequency light with a time dependent amplitude can be used to
parametrically excite and amplify THz plasmons, mirror plasmonic wave
packets in time, generate momemtum-gapped plasmonic band structures,
entangled plasmon pairs, and THz radiation in two dimensional Dirac
systems. Our results show how low frequency plasmons can be coherently
excited and manipulated without the need for THz light.
\end{abstract}
\maketitle

\paragraph*{Introduction.}

Time varying media \citep{galiffi2022photonics_time_varying_ptc},
discussed by Morgenthaler as early as 1958 \citep{morgenthaler1958velocity_modulated},
and by Holberg and Kunz in 1966 \citep{holberg1966parametric_PTC},
have recently attracted a great deal of attention due to a variety
of exotic effects and potential applications. Notable examples are
time-varying mirrors \citep{tirole2022_time_varying_mirror}, space-time
crystals \citep{peng2022topological_spacetime,sharabi2022spatiotemporal_PTC},
time-varying metasurfaces \citep{wang2023metasurface_PTC,wang2023controlling_surface_waves,koutserimpas2023multiharmonic_time_varied_metasurfaces},
temporal switching \citep{akbarzadeh2018_temporal_switch,pacheco2020temporal_switching,pacheco2021temporal_switching2},
amplified emission and lasing \citep{lyubarov2022photonic_time_crystal_amplified},
photon pair generation \citep{mendoncca2005_entangled_photon_pairs,lyubarov2022photonic_time_crystal_amplified},
nontrivial topological \citep{lustig2018topological_photonic_time_crystal}
and statistical properties \citep{carminati2021universal_statistics_time-varying_medium},
non-Hermiticity \citep{wang2018photonic_non_hermitian,li2021_non_hermition_TPT},
momentum-gapped (k-gapped) states, and unusual solitonic behavior
\citep{pan2022superluminal_k-gap_solitons}. Beyond photonics, time
varying media have been realized in classical liquids \citep{bacot2016time_water}
and acoustic media \citep{fleury2016_topological_sound,wen2022_acoustic_non_hermitian}.

Floquet engineering of quantum materials, on the other hand, is a
powerful tool for the manipulation of band structures and for creating
non-equilibriun correlated states in atomic, optical, and condensed
matter physics \citep{oka2009photovoltaic_hall_effect,kitagawa2011floquetinduced,kitagawa2010topological_characterization_driven_quantum_system,lindner2011floquet,wang2013_floquet-bloch_states_observation,mciver2020light_anomaouls_hall_graphene,mahmood2016selective_scattering_floquet-bloch_volkov,zhou2023black_phosphorus_floquet,usaj2014_floquet_graphene_topo,perez2014floquet_traphene_topo,oka2019floquet_review,katz2020optically,castro2022optimal_floquet_control,esin2018q_steady_state_topo_ins,esin2020floquet_metal_insulator,esin2021_liquid_crystal,dehghani2015_floquet_topo,genske2015floquet_boltzmann,glazman1983kinetics_pulses_semiconductor,dehghani2014dissipative_topo_floquet,sentef2015pump_probe_floquet,chan2016floquet_ref,farrell2015floquet_ref,gu2011floquet_ref,hubener2017floquet_ref,jiang2011floquet_ref,kennes2019floquet_ref,kundu2013floquet_ref,thakurathi2017floquet_ref}.
A recent work suggested that Floquet engineering using high frequency
drives enables control of low-energy collective modes of many body
systems through Modulated Floquet Parametric Driving (MFPD) and leads
to new correlated states \citep{kiselev2023MFPD}. Here, we demonstrate
that an amplitude modulated optical Floquet drive can be used to create
momentum gapped plasmonic time varying media -- teraherz (THz) analogues
of so called photonic time crystals (PTCs) \citep{lustig2018topological_photonic_time_crystal}.
Plasmonic time varying media can be used to create entangled plasmon
pairs, amplify plasmons, as well as reverse their propagation in time.
MFPD surpasses experimental difficulties associated with the excitation
and manipulation of low frequency plasmons: A high frequency ($\mathrm{eV}$-range)
drive is used to engineer an effective band structure. Varying the
drive's amplitude in time then creates a medium with time dependent
properties.

\begin{figure}[h]
\centering{}\includegraphics[width=1\columnwidth]{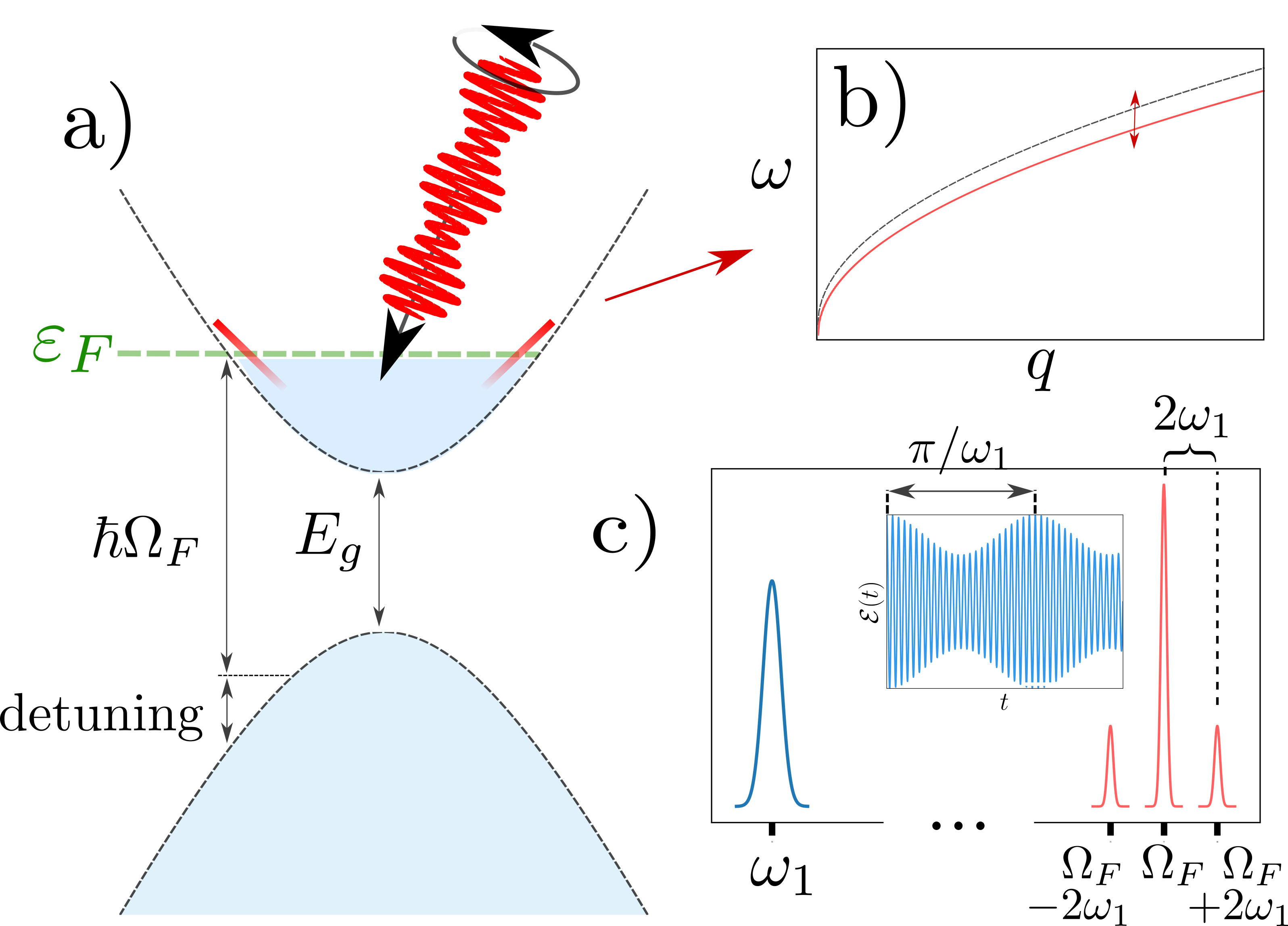}\caption{Floquet engineering and MFPD. a) The band structure (grey) of a gapped
Dirac material, e.g. a transition metal dichalcogenide, with gap $E_{g}$
driven with circularly polarized infrared light. The driving frequency
$\Omega_{F}$ and the Fermi energy $\varepsilon_{F}$ are chosen detuned,
such that single photon absorption and particle-hole creation with
energy transfer $\hbar\Omega_{F}$ is blocked by the Fermi see. The
driving leads to a change of the group velocity at the Fermi surface
(red). b) This change results in a shift of the plasmon dispersion
$\omega_{\mathrm{pl}}\left(q\right)$. c) Modulating the amplitude
of the Floquet drive with a frequency $\omega_{1}\ll\Omega_{F}$ will
lead to an oscillating Fermi velocity. This oscillation induces a
parametric resonance of the plasmon modes. The spectrum of the modulated
signal consists of a strong central peak at $\Omega_{F}$ and two
side-bands at $\Omega_{F}\pm2\omega_{1}$ (red peaks). The parametric
generation of plasmons at half the difference frequency (blue peak)
corresponds to frequency mixing.\label{fig:floquet_engineering}}
\end{figure}

Compared to many other solid state Floquet engineering schemes, MFPD
offers two major advantages: First, the Floquet drive couples to a
resonant mode (the plasmon), which stores its energy over many oscillation
cycles and thus amplifies its effects. Second, it operates in a regime
with strongly suppressed photon absorption and heating (see Fig. \ref{fig:floquet_engineering}).

While we focus on THz plasmons in this manuscript, we note that the
MFPD principle is very general, and could be employed to create time
varying media with other kinds of collective modes and at different
frequencies. The main limitation with respect to the frequency range
is that the amplitude modulation, which is responsible for the time-varying
properties, is significantly slower than the carrier frequency of
the Floquet drive. Potentially, MFPD can be applied to magnons in
ferro- and antiferromagnets, which span a range of frequencies between
GHz and THz \citep{rezende2019introduction_antiferromagnetic_magnons,kreisel2009_YIG_magnon_spectra}.
Here similar effects could be mediated by the renormalization of the
effective exchange constant by coherent light \citep{mentink2015_Mott_exchange_floquet,chaudhary2019orbital_floquet_engineering_exchange,ron2020_light_induced_enhancement_of_exchange}.

Our starting point are electrons in a two dimensional material. An
effective quasi-energy band structure is induced by driving the electrons
coherently with light of frequency $\Omega_{F}$. The induced band
structure depends on the amplitude of the driving signal. A subsequent
periodic modulation of the drive amplitude with a frequency $2\omega_{1}\ll\Omega_{F}$
results in a periodically changing Fermi velocity. The oscillation
of the Fermi velocity parametrically couples to the soft plasmon modes
of the two dimensional electron gas. This is the principle of MFPD,
which is illustrated in Fig.\ref{fig:floquet_engineering}. In this
letter, we will consider plasmons with low frequencies of a few THz
\citep{lundeberg2017_koppens_plasmonics_near_field}. The principle,
however, can be applied more broadly, depending on material and driving
parameters.

In the linear approximation, the Floquet-engineered dynamics of the
plane wave plasmon modes $\delta\rho_{\mathbf{q}}$ is governed by
the equation
\begin{equation}
\partial_{t}\left(1+h\cos2\omega_{1}t\right)\partial_{t}\delta\rho_{\mathbf{q}}+\omega_{\mathrm{pl}}^{2}\left(q\right)\delta\rho{}_{\mathbf{q}}=0,\label{eq:time_varying_medium_eq}
\end{equation}
which is derived below. Here $h$ is an amplitude describing the effect
of a slow modulation of the driving field (see Eq. (\ref{eq:Oscillating_mass})),
and $\omega_{\mathrm{pl}}\left(q\right)$ is the plasmon dispersion
defined in Eq. (\ref{eq:equilib_plasmon_modes}). Equation (\ref{eq:time_varying_medium_eq})
is a plasmonic version of the equation describing the evolution of
electromagnetic waves in PTCs \citep{lyubarov2022photonic_time_crystal_amplified},
where the driving enters through a modulation of the dielectric constant
$\varepsilon\left(t\right)$.

In essence, Eq. (\ref{eq:time_varying_medium_eq}) is a parametric
oscillator equation describing parametrically excited plasmonic plane-wave
modes. We note that momentum conservation requires that modes with
wavevectors $\mathbf{q}$ and $-\mathbf{q}$ are equally excited,
such that the net momentum is zero \citep{kiselev2023MFPD}. Parametric
resonance occurs for $\omega_{\mathrm{pl}}\left(q\right)\approx n\omega_{1}$,
with $n=1,2,3...$ (see Ref. \citep{landau_lifshitz_mechanics} §
27). In the presence of damping, resonance occurs above a critical
amplitude $h_{c}$, which grows with increasing $n$ \citep{turyn1993_mathieu_threshold_high_m}.
We will focus on the $n=1$ case, which requires the lowest $h_{c}$.
It can be useful to exploit resonances with $n>1$ to lower the required
modulation frequency \citep{zurita2009reflection_epsilon_time_dep,asadchy2022_time_dependent_scatterers},
however modulation frequencies of up to $10\:\mathrm{THz}$ can be
reached by interfering laser beams with slightly detuned frequencies
\citep{rubinsztein2016roadmap_structured_light,wheaton2015_modulation_EAR}.

This reveals another basic functionality of MFPD. It can be used as
a frequency mixer, which creates standing wave plasmons at frequencies
corresponding to the beating of the high frequency signal (see Fig.
\ref{fig:floquet_engineering}c)). These plasmons, which will have
$\upmu\mathrm{m}$-wavelengths and create strong electric fields in
the THz range (see the discussion of experimental parameters at the
end of this letter), can be detected by near field optical microscopy
\citep{chen2012optical_near_field_tip_plasmons_scanning_koppens,fei2012_near_field_tip_plasmons_scanning_basov}.

Interesting effects on plasmon propagation can also be achieved with
pulsed signals. In analogy to recent experiments with surface waves
in classical liquids \citep{bacot2016time_water}, we propose that
pulses can be used to induce time reversal of propagating plasmon
wave packets via a temporary change of the electrons' effective mass.
A high frequency pulse will split a propagating plasmon wave packet
into two parts. While one part will continue to propagate in the initial
direction, the other will evolve backwards in time and propagate towards
the origin. We suggest that pulse induced time reversal is a promising
way to control the propagation of THz plasmons with off-resonant,
high frequency light, and discuss this functionality in more detail
towards the end of this letter.

\paragraph{Modulated Floquet parametric driving: Floquet engineering of time
varying materials.}

We now show, in more detail, how MFPD can be used to create a time
varying plasmonic medium. Consider a gapped Dirac Hamiltonian which
describes electrons near the Fermi level of a two dimensional material
-- e.g. a transition metal dichalcogenide or black phosphorus \citep{chaves2020_2d_semiconductors_bandgaps,kim2015dirac_black_phosphorus,chaves2017_excitonic_tmdcs}:

\begin{equation}
H=\sum_{\mathbf{k}}\mathbf{c}_{\mathbf{k}}^{\dagger}\left[H_{0}\left(\mathbf{k}\right)+H_{d}\left(t\right)\right]\mathbf{c}_{\mathbf{k}}+\sum_{\mathbf{q}}V\left(\mathbf{q}\right)\rho_{\mathbf{q}}\rho_{-\mathbf{q}}.\label{eq:general_H}
\end{equation}
We focus on a single valley and write $H_{0}=\mathbf{d}\cdot\boldsymbol{\sigma}$
with $\mathbf{d}=\left[\lambda k_{x},\lambda k_{y},E_{g}/2\right]$,
where, $E_{g}$ is the energy gap between the two bands, $\boldsymbol{\sigma}$
is a Pauli vector describing pseudospin-orbit coupling, $\mathbf{c}_{\mathbf{k}}^{\dagger}$,$\mathbf{c}_{\mathbf{k}}$
are electron creation and annihilation operators, $\rho_{\mathbf{q}}=\sum_{\mathbf{k}}\mathbf{c}_{\mathbf{k}+\mathbf{q}}^{\dagger}\mathbf{c}_{\mathbf{k}}$
is the density operator and $V\left(\mathbf{q}\right)=2\pi/q$ is
the 2D Fourier transform of the Coulomb potential. The driving part
of the Hamiltonian is derived from minimal coupling: $H_{d}\left(\mathbf{k}\right)=e\mathbf{A}\cdot\nabla_{\mathbf{k}}H_{0}$.
We assume driving by circularly polarized light with an amplitude
$\mathcal{E}$ described by the vector potential $\mathbf{A}=\left(\mathcal{E}/\Omega_{F}\right)\left[-\sin\Omega_{F}t,\cos\Omega_{F}t,0\right]$.
Spin-valley degeneracy is included in the final results. 

To avoid direct single photon absorptions, which are a dominant source
of heating in Floquet engineered systems \citep{seetharam2015baths_controlled_floquet_population,esin2018q_steady_state_topo_ins,esin2021_liquid_crystal},
we suggest to operate in an off-resonant regime, where the Fermi surface
lies close to, but above the single photon resonance (see Fig.\ref{fig:floquet_engineering}).
In this off-resonant regime, all states supporting excitations of
a single electron by photons with an energy of $\hbar\Omega_{F}$
and near-zero momentum transfer are blocked. Processes involving the
absorption of multiple photons are suppressed to second order in the
small ratio of driving amplitude over driving frequency $e\mathcal{E}\lambda/\Omega_{F}^{2}\hbar^{2}$\citep{seetharam2015baths_controlled_floquet_population,esin2021_liquid_crystal}.
Interaction and disorder assisted single photon absorption is possible,
but can be shown to not cause an overheating of the system (see Supplement
G).

Let us now carry out the MFPD program and subject the Floquet drive
amplitude $\mathcal{E}$ to a slow modulation. Expanding the dispersion
relation around the Fermi momentum \footnote{Note that $k_{F}$ is fixed by momentum conservation.}
we find $\varepsilon_{k}\approx\hbar v_{F}\left(\mathcal{E},\Omega_{F}\right)\left(k-k_{F}\right)+\varepsilon_{k_{F}}\left(\mathcal{E},\Omega_{F}\right)$
\citep{abrikosov1959}. Here, the Fermi velocity $v_{F}\left(\mathcal{E},\Omega_{F}\right)$
depends on the amplitude and frequency of the Floquet drive. A slow,
abiabatic, time periodic modulation of $\mathcal{E}$ according to
$\mathcal{E}\left(t\right)=\bar{\mathcal{E}}+\delta\mathcal{E}\cos\left(2\omega_{1}t\right),$where
$\omega_{1}\ll\Omega_{F}$, will result in an oscillating Fermi velocity.
For our purposes, it will be convenient to parametrize this time dependence
as an oscillation of the effective mass $m^{*}=\hbar k_{F}/v_{F}\left(\mathcal{E},\Omega_{F}\right)$:
\begin{equation}
m^{*}\left(t\right)=\bar{m}^{*}\left(1+h\cos\left(2\omega_{1}t\right)\right).\label{eq:Oscillating_mass}
\end{equation}
Here, $h$ is a small dimensionless number quantifying the amplitude
of the oscillatory component of the effective mass. Below, we estimate
that for reasonable driving strengths $h$ is of the order of $10^{-2}$.
Explicit formulas for $\varepsilon_{k}\left(\mathcal{E},\Omega_{F}\right)$,
$v_{F}\left(\mathcal{E},\Omega_{F}\right)$ and $\bar{m}^{*}\left(\mathcal{E},\Omega_{F}\right)$
are given in Supplement A.

\paragraph{Momentum-gapped states.}

We now investigate the influence of the oscillating effective mass
of Eq. (\ref{eq:Oscillating_mass}) on the dispersion relation of
plasmons in a Coulomb interacting 2D electron gas. The plasmon dynamics
can be inferred from charge and momentum conservation \citep{Eguiluz1976hydrodynamicPlasmons,Forster,lucas2015memory,kiselev2021_superdiffusive_modes}.
An inhomogeneous charge distribution $\rho\left(\mathbf{x},t\right)$
will induce an electrostatic potential $\phi\left(\mathbf{x},t\right)$
in the system, which will accelerate the electrons according to Newton's
law: $\partial_{t}\mathbf{p}\left(\mathbf{x},t\right)=-e\rho\left(\mathbf{x},t\right)\nabla\phi\left(\mathbf{x},t\right)$.
Here, $\mathbf{p}$ is the momentum density of the electrons. Combining
the continuity equation $\partial_{t}\rho\left(\mathbf{x},t\right)=-\nabla\cdot\mathbf{j}\left(\mathbf{x},t\right)$
and the relationship between momentum and current densities $e\mathbf{p}\left(\mathbf{x},t\right)=m^{*}\left(t\right)\mathbf{j}\left(\mathbf{x},t\right)$
(see Supplement B) with Newton's law, we find
\begin{equation}
\partial_{t}m^{*}\left(t\right)\partial_{t}\rho\left(\mathbf{x},t\right)=\frac{e}{4\pi\varepsilon}\nabla\cdot\rho\left(\mathbf{x},t\right)\nabla\int d^{2}x'\frac{\rho\left(\mathbf{x}',t\right)}{\left|\mathbf{x}-\mathbf{x}'\right|},\label{eq:dyn_real_space}
\end{equation}
where the integral extends over the sample. We separate the oscillating
part of $\rho$ from the homogeneous background $\bar{\rho}$ and
write $\rho=\bar{\rho}+\delta\rho$. Linearizing Eq. (\ref{eq:dyn_real_space})
in $\delta\rho$, and assuming plane-wave solutions $\delta\rho\left(\mathbf{x}\right)=\delta\rho_{\mathbf{q}}\exp\left(i\mathbf{q}\cdot\mathbf{x}\right)$,
we obtain Eq. (\ref{eq:time_varying_medium_eq}).

The plasmon dispersion $\omega_{\mathrm{pl}}\left(q\right)$ is given
by
\begin{equation}
\omega_{\mathrm{pl}}\left(q\right)=\sqrt{\frac{e\bar{\rho}}{4\pi\bar{m}^{*}\varepsilon}q^{2}V\left(q\right)}.\label{eq:equilib_plasmon_modes}
\end{equation}
$V\left(q\right)$ is the Fourier transform of the Coulomb potential.

Consider a quasi one-dimensional, waveguide-like set-up of infinite
length and width $l$, oriented parallel to the $x$-axis. For this
strip geometry we find
\begin{equation}
V\left(q\right)\approx\int_{0}^{l}dy\int_{-\infty}^{\infty}dx\frac{e^{i\mathbf{q}\cdot\mathbf{x}}}{\left|\mathbf{x}\right|}\approx2l\left|\ln\frac{\left|q\right|l}{4}\right|\label{eq:quasi-1d-pot}
\end{equation}
for $q\ll1/l$, where $\mathbf{q}=q\hat{\mathbf{e}}_{x}$. The precise
geometry of the device is not very important for the physics presented
here. The calculations would be very similar for an infinite 2D sample
with $V\left(q\right)=2\pi/q$, but we would need to keep track of
the vector nature of $\mathbf{q}$.

Let us study the behavior of $\delta\rho_{q}$ in the vicinity of
the parametric resonance, where $\left|\omega_{\mathrm{pl}}\left(q\right)-\omega_{1}\right|\approx h\omega_{1}$.
Using the ansatz $\delta\rho_{q}=a_{q}\left(t\right)\cos\left(\omega_{1}t\right)+b_{q}\left(t\right)\sin\left(\omega_{1}t\right)$,
where the slowly varying coefficients are given by $a_{q}\left(t\right)=\tilde{a}e^{st}$
and $b_{q}\left(t\right)=\tilde{b}e^{st}$, we find (see Supplement
C)
\begin{equation}
s_{\pm}\left(q\right)=\pm\frac{\omega_{1}}{2}\sqrt{\frac{1}{4}h^{2}-\left(\frac{\omega_{\mathrm{pl}}^{2}\left(q\right)-\omega_{1}^{2}}{\omega_{1}^{2}}\right)^{2}}.\label{eq:gap_bands}
\end{equation}
For $\left|\omega_{\mathrm{pl}}^{2}\left(q\right)-\omega_{1}^{2}\right|>\omega_{1}^{2}h/2$
the exponent is imaginary, corresponding to dispersive plane-wave
solutions, whereas for $\left|\omega_{\mathrm{pl}}^{2}\left(q\right)-\omega_{1}^{2}\right|<\omega_{1}^{2}h/2$,
we find exponentially growing (or decaying) unstable non-dispersive
solutions oscillating at the frequency $\omega_{1}$. This results
in a momentum gapped dispersion, similar to the one appearing in so
called photonic time crystals \citep{lustig2018topological_photonic_time_crystal,galiffi2022photonics_time_varying_ptc}.
The gap is centered at the wavevectors $\pm q^{*}$, where $q^{*}$
is determined by the resonance condition $\omega_{\mathrm{pl}}\left(q^{*}\right)=\omega_{1}$.

\begin{figure}
\begin{centering}
\includegraphics[width=0.95\columnwidth]{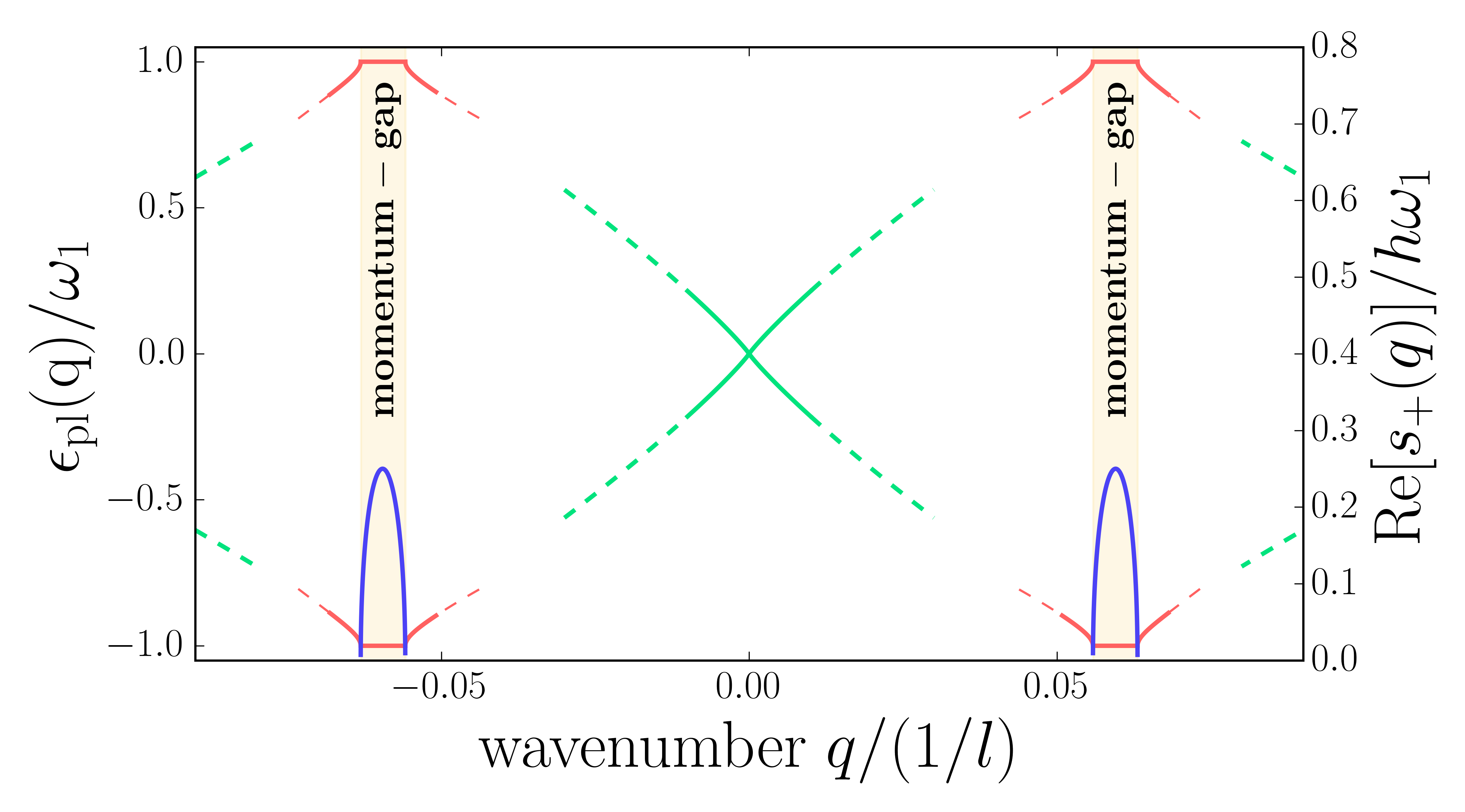}
\par\end{centering}
\centering{}\caption{Dispersion of the plasmonic time varying medium. The modulated Floquet
drive couples to plasmon modes and induces gaps (yellow segments)
near the resonance $\omega_{\mathrm{pl}}\left(q\right)=\pm\omega_{1}$.
In the vicinity of the gaps, the modes are given by Eq. (\ref{eq:gap_bands})
(red curves). Away from the gap, the drive-induced changes to the
dispersion are small (green curves) and Eqs. (\ref{eq:equilib_plasmon_modes})
and (\ref{eq:quasi-1d-pot}) are approximately valid. The growth rate
of the unstable modes $\mathrm{Re}\left[s_{+}\left(q\right)\right]$
is shown in blue. \label{fig:q_gaps}}
\end{figure}

Notice that Eq. (\ref{eq:time_varying_medium_eq}) is invariant with
respect to time translations by $T=\pi/\omega_{1}$. This symmetry
is broken by the system's response -- a feature that is known from
discrete time crystals \citep{else2020discrete_time_cryst_review,else2016floquet_discrete_time_crystals,yao2017discrete_time_crystal_original,zhang2017observation_discrete_time_crystal,kyprianidis2021observation_discrete_time_crystal,natsheh2021critical_properties_time_crystal,yao2020classical_discrete_time_crystal}
and generic to parametric resonances. The momentum gap and the dispersion
in the vicinity of the gap are shown in Fig. \ref{fig:q_gaps}.

So far we did not consider the finite lifetime of plasmons. For small
$q$, the plasmon dispersion lies outside the particle-hole continuum
and Landau damping can be neglected (see Supplement G). In clean materials
the main source of damping is momentum relaxing phonon scattering
\citep{ni2018_plasmon_quality_factor_graphene}. To model this effect,
we add momentum decay at a rate $\gamma$ to Newton's law $\left(\partial_{t}+\gamma\right)\mathbf{p}\left(\mathbf{x},t\right)=-\rho\left(\mathbf{x},t\right)\nabla\phi\left(\mathbf{x},t\right)$.
To first order in $\gamma$, $s_{\pm}\left(q\right)$ in Eq. (\ref{eq:gap_bands})
obtains a negative real part of $-\gamma/2$ (see Supplement C). The
instability condition $\mathrm{Re}\left(s_{\pm}\right)>0$ is realized
in a narrow region around $q^{*}$ for 
\begin{equation}
h>2\gamma/\omega_{1}.\label{eq:instability_damping_condition}
\end{equation}
Interestingly, even if $h<2\gamma/\omega_{1}$ holds, the momentum-gap
remains intact. The gap then hosts non-dispersive modes with decay
rates $-\gamma/2\pm\omega_{1}h^{2}/4$ at $q=q^{*}$.

Eq. (\ref{eq:gap_bands}) suggests that the opening of the momentum
gap is associated with the joining of two branches of the plasmon
dispersion described by the plus and minus signs. To interpret the
two branches, it is useful to write the solution of Eq. (\ref{eq:gap_bands})
in the form known from Floquet's theorem \citep{ince1956o_ODE_book}:
\begin{equation}
\delta\rho_{q}=e^{-i\epsilon_{\mathrm{pl}}\left(q\right)t}u_{\epsilon_{\mathrm{pl}}}\left(t\right),
\end{equation}
where $\epsilon_{\mathrm{pl}}\left(q\right)$ is the plasmon quasi-dispersion
that determines wave propagation once the periodic modulation is switched
on \footnote{The quantity $\epsilon_{\mathrm{pl}}\left(q\right)$ is analogous
to the quasi-energy of floquet driven electrons.}, and $u_{\epsilon_{\mathrm{pl}}}\left(t\right)=e^{-i\epsilon_{\mathrm{pl}}t}\left[a\left(e^{2i\left(\epsilon_{\mathrm{pl},\pm}+is_{\pm}\right)t}+1\right)\pm ib\left(e^{2i\left(\epsilon_{\mathrm{pl},\pm}+is_{\pm}\right)t}-1\right)\right]/2$,
where the plus is chosen for $\epsilon_{\mathrm{pl}}<0$ and vice
versa. It is easy to verify that $u_{\epsilon_{\mathrm{pl}}}\left(t+\pi/\omega_{1}\right)=u_{\epsilon_{\mathrm{pl}}}\left(t\right)$,
in accordance with Floquet's theorem. The quasi-dispersion $\epsilon_{\mathrm{pl}}\left(q\right)$
can be confined to a Brillouin zone of width $2\omega_{1}$. We choose
$\pm\omega_{1}$ as the boundaries of the Brillouin zone. Away from
the gap, for small $q$, the dispersion is given by $\epsilon_{\mathrm{pl}}\left(q\right)\approx\pm i\omega_{\mathrm{pl}}\left(q\right)/\sqrt{1+h^{2}/4}$,
showing that the impact of the drive is small. The plasmonic band
structure in terms of $\epsilon_{\mathrm{pl}}$ is shown in Fig. \ref{fig:q_gaps}.

\paragraph{Production of entangled plasmon pairs.}

Since the momentum supplied by the optical drive is negligible, plasmons
can only be produced in pairs with opposite momenta $\pm\hbar q$.
This results in generation of entangled plasmon pairs. To see this,
we find the Hamiltonian leading to Eq. (\ref{eq:time_varying_medium_eq})
and then apply a second quantization procedure following Refs. \citep{mendoncca2005_entangled_photon_pairs,lyubarov2022photonic_time_crystal_amplified}.
Details are given in Supplement D. For the driven, resonant plasmon
modes we find the Hamiltonian in the interaction representation
\begin{equation}
H_{\mathrm{pl},\mathrm{int}}\approx\frac{h}{4}\hbar\omega_{1}\left(a_{q^{*}}^{\dagger}a_{-q^{*}}^{\dagger}+a_{-q^{*}}a_{q^{*}}\right),\label{eq:Interaction_h_gen_plasmons}
\end{equation}
where $a_{q^{*}}^{\dagger}$ and $a_{q^{*}}$ are plasmon creation
and annihilation operators for the resonant wavenumber $q^{*}$.

Applied to the vacuum, the time evolution operator $U_{\mathrm{pl,int}}\left(t,0\right)=\exp\left(-tH_{\mathrm{pl},\mathrm{int}}/\hbar\right)$
generates a non-factorizable two mode squeezed state $\ket{\psi_{\mathrm{pl}}\left(t\right)}=U_{\mathrm{pl,int}}\left(t,0\right)\ket{0,0}$
in the basis $\ket{n_{q},n_{-q}}$ \citep{gerry2005introductory_quantum_optics},
where $n_{q}$ gives the number of plasmons in the state $q$. The
generation of entangled plasmon pairs has been discussed previously
in Ref. \citep{sun2022graphene_entangled_plasmon_pairs}. The authors
suggested to excite the longitudinal plasmon modes of a graphene ribbon
via their coupling to a resonantly pumped transverse mode and discussed
modes of detection.

\paragraph{Pulse induced time reversal}

\begin{figure}[h]
\centering{}\includegraphics[width=1\columnwidth]{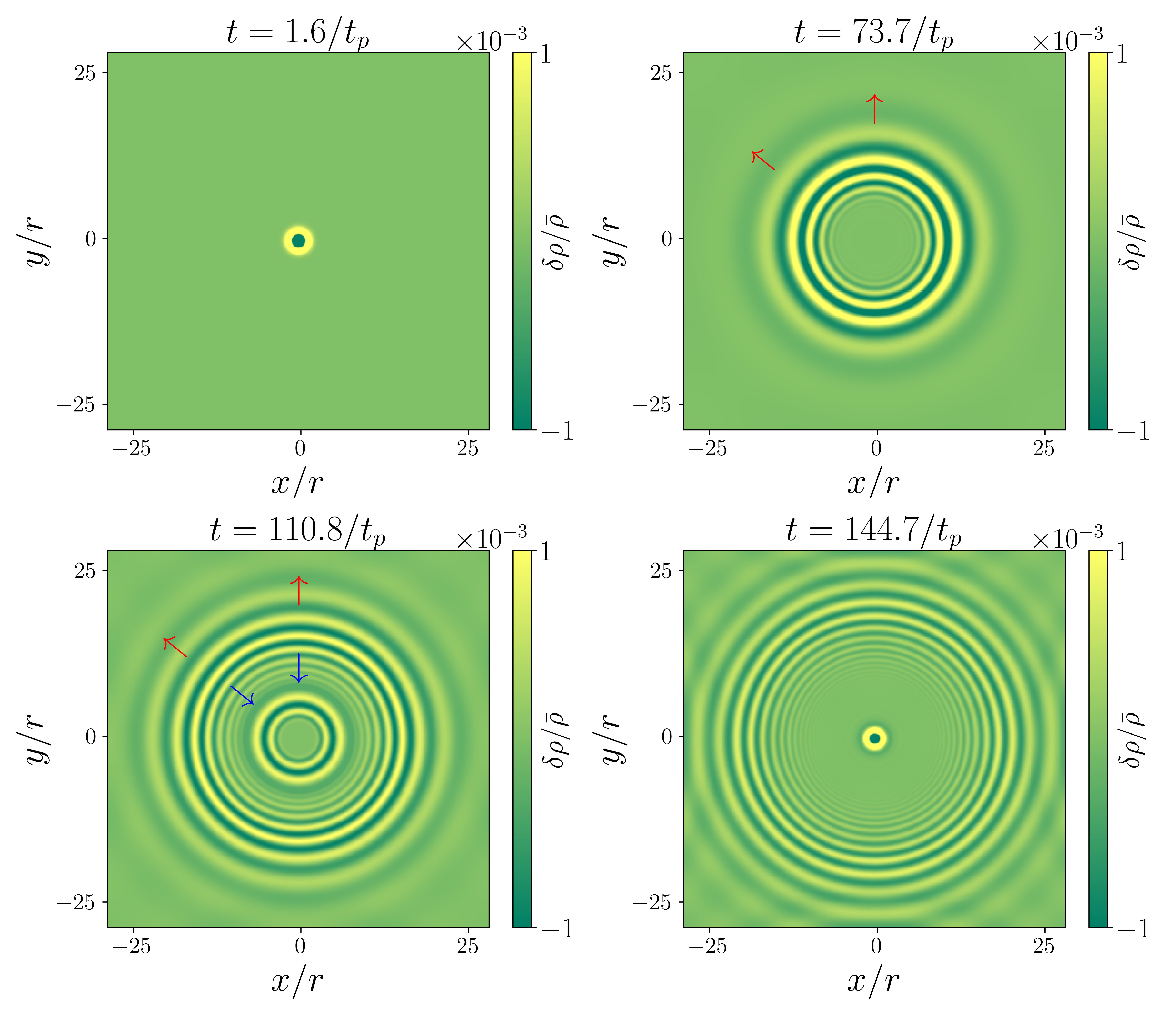}\caption{Numerical simulation of pulse induced time reversal of a propagating
cylindical plasmon wave packet. The plasmon is triggered by an electromagnetic
pulse with duration $t_{p}$ localized in a region with radius $r$
around the origin. A high frequency gaussian pulse, resulting in a
change of the electron's Fermi velocity and effective mass, is applied
at time $t_{0}=75t_{p}$. The wavefront is splitted into parts, with
the time-reversed part propagating back towards the origin (blue arrows).
At $2t_{0}$, the time reversed front is focused back at the origin.
For this simulation, Eq. (\ref{eq:dyn_real_space}) was solved using
the Dedalus package \citep{burns2020dedalus}. \label{fig:time_reversal}}
\end{figure}

So far we have considered the action of a periodic oscillation of
the effective mass $m^{*}\left(t\right)$ resulting from an amplitude
modulated high frequency Floquet driving. We now discuss how a pulse
induced, abrupt change of $m^{*}\left(t\right)$ can time reverse
a propagating plasmon wave packet. The effect is shown in Fig.\ref{fig:time_reversal}.
We consider a pulse with peak amplitude $\mathcal{E}_{0}$, center
frequency $\Omega_{F}$ and duration $\Delta t$. As shown in Supplement
E, the effect of such a pulse at time $t=t_{0}$ on the effective
mass can be described by the formula $m^{*}\left(t\right)=\bar{m}^{*}\left(1+\bar{h}\delta\left(t-t_{0}\right)\right)$,
where $\bar{h}$ is an amplitude characterizing the impact of the
pulse. This assumes, that the pulse is very short compared to a plasmon
oscillation cycle. For an incident wave packet $\delta\rho_{i}\left(\mathbf{x},t\right)$,
the time reversed part, which appears after the pulse, is given by
(Supplement E)
\begin{equation}
\delta\rho_{\mathrm{rev}}\left(\mathbf{x},t\right)\approx A\left(q_{0}\right)\delta\rho_{i}\left(\mathbf{x},-t\right).
\end{equation}
Here, $A\left(q_{0}\right)=\frac{1}{2}\omega_{\mathrm{pl}}\left(q_{0}\right)\bar{h}e^{-i2\omega_{\mathrm{pl}}\left(q_{0}\right)t_{0}}$
is a complex amplitude and $q_{0}$ is the central wavenumber of the
packet.

The assumption of a delta-function like, abrupt change of $m^{*}$,
while convenient for analytic calculations (Supplement E), is only
valid if the pulse duration is much shorter than the time scale of
the plasmon oscillations. To achieve a strong reshaping of the electron
dispersion, typically pulses with field strenghts of $10^{7}-10^{8}\,\mathrm{V}/\mathrm{m}$
and durations of around $100\,\mathrm{fs}$ are employed \citep{zhou2023black_phosphorus_floquet,wang2013_floquet-bloch_states_observation}.
Being interested in the pulse induced time reversal of plasmons with
frequencies of a few THz, we address the question of whether pulses
with durations of up to half the plasmon oscillation cycle $T_{\mathrm{pl}}$
can be used. Using the Dedalus library \citep{burns2020dedalus},
we carry out simulations (Supplementary E) for pulse durations of
$\Delta t=0.1T_{\mathrm{pl}}$ (see Fig. \ref{fig:time_reversal}),
$\Delta t=0.22T_{\mathrm{pl}}$, and $\Delta t=0.5T_{\mathrm{pl}}$.
We find that pulse induced time reversal is observable for all three
values.

\paragraph{Discussion.}

The reshaping of bandstructure by light in the off-resonant regime
considered here has recently been demonstrated in doped black phosphorus
\citep{zhou2023black_phosphorus_floquet}, showing that the regime
of operation for MFPD controlled plasmonic time varying materials
is within experimental reach. In Supplement F, we estimate the critical
driving strength to be $\mathcal{\bar{E}}\approx4\cdot10^{5}\,\mathrm{V}/\mathrm{m}$
using Eq. (\ref{eq:instability_damping_condition}) and the parameters
$\bar{\rho}=1.18\cdot10^{11}/\mathrm{cm}^{-2}$, $E_{g}=0.3\,\mathrm{eV}$,
$\hbar\Omega_{F}=0.35\,\mathrm{eV}$, $\lambda=15\,\mathrm{eV}\text{Å},$$\varepsilon=6\varepsilon_{0}$.
The corresponding laser intensity is orders of magnitude smaller then
in current solid state Floquet experiments \citep{wang2013_floquet-bloch_states_observation,mahmood2016selective_scattering_floquet-bloch_volkov,mciver2020light_anomaouls_hall_graphene,zhou2023black_phosphorus_floquet}.
The choice of material parameters was inspired by black phosphorus,
and can be adjusted for other materials. The estimate assumes a plasmon
quality factor of $Q=\omega_{1}/\gamma\approx10^{2}$, reachable in
clean materials \citep{ni2018_plasmon_quality_factor_graphene,kitagawa2011floquetinduced,mciver2020light_anomaouls_hall_graphene}.
In theory, $Q$-factors of $10^{3}-10^{4}$ are believed to be achievable
\citep{principi2013intrinsic_graphene_plasmon_quality_factor,ni2018_plasmon_quality_factor_graphene},
which would further reduce the necessary driving power. Moreover we
demonstrate in Supplement G that the heating caused by photon absorption
from MFPD can be compensated for by the cooling power of the lattice.
We stress that the realization of plasmonic time reversal requires
only pulsed signals and could be demonstrated with set-ups similar
to the ones used in most Floquet experiments \citep{mciver2020light_anomaouls_hall_graphene,zhou2023black_phosphorus_floquet,mahmood2016selective_scattering_floquet-bloch_volkov,wang2013_floquet-bloch_states_observation}.
Near-field optical microscopy can be used for detection.

To estimate the experimental parameters, we consider plasmons with
frequencies of $1-5\,\mathrm{THz}$, and use the above values for
material and driving parameters. Table \ref{tab:exp_param_table}
shows our estimates for plasmon wavelengths and field strengths. The
momentum gap has typically a width of $1-2\%$ of $q^{*}$, e.g. $0.05/\upmu\mathrm{m}$
for the strip and $1\,\mathrm{THz}$. To estimate the field strengths,
we draw on Ref. \citep{kiselev2023MFPD}, which predicts that, due
to nonlinear effects, the exponential plasmon growth saturates at
amplitudes of $\sim\sqrt{8h_{c}}\bar{\rho}\epsilon^{1/4}$, where
$\epsilon$ is the relative distance of the drive from the critical
driving strength. We assume a driving that is $10\%$ above threshold,
giving $\epsilon=0.1$.

\begin{table}
\centering{}%
\begin{tabular}{|c|c|c|c|}
\hline 
sample & frequency & wavelength & field strength\tabularnewline
\hline 
\hline 
2D & 1 THz & $8.2\,\upmu\mathrm{m}$ & $2.3\cdot10^{5}\,\mathrm{V}/\mathrm{m}$\tabularnewline
\hline 
2D & 5 THz & $0.3\,\upmu\mathrm{m}$ & $2.3\cdot10^{5}\,\mathrm{V}/\mathrm{m}$\tabularnewline
\hline 
strip, $l=0.2\,\upmu\mathrm{m}$ & 1 THz & $1.7\,\upmu\mathrm{m}$ & $1.6\cdot10^{5}\,\mathrm{V}/\mathrm{m}$\tabularnewline
\hline 
strip, $l=0.05\,\upmu\mathrm{m}$ & 3 THz & $0.2\,\upmu\mathrm{m}$ & $2.8\cdot10^{5}\,\mathrm{V}/\mathrm{m}$\tabularnewline
\hline 
\end{tabular}\caption{Plasmon frequencies, wavelengths and field strengths of MFPD driven
plasmons for a two dimensional and strip-like samples. \label{tab:exp_param_table}}
\end{table}
Summing up, we proposed a method to induce low frequency plasmons
by amplitude modulated high frequenecy signals, and to create time
varying materials for plasmons with effects such as parametric amplification,
the opening of momentum gaps, time reversal mirroring, and creation
of entangled plasmon pairs. The proposed effects can be measured with
near field microscopy \citep{fei2012_near_field_tip_plasmons_scanning_basov,chen2012optical_near_field_tip_plasmons_scanning_koppens,lundeberg2017_koppens_plasmonics_near_field}
and used in divice applications, e.g. as plasmon sources or to control
the propagation of plasmons.
\begin{acknowledgments}
We acknowledge useful conversations with D. Basov and M. Rudner. E.K.
thanks the Helen Diller quantum center for financial support.
\end{acknowledgments}


\clearpage\onecolumngrid

\setcounter{page}{1}
\section*{Supplementary Material} 

\renewcommand{\thesection}{Supplementary Sec.}%

\setcounter{figure}{0}
\renewcommand{\thefigure}{S\ \arabic{figure}}%
\renewcommand{\figurename}{Supplementary Figure}

\setcounter{equation}{0}
\renewcommand{\theequation}{S\,\arabic{equation}}%

\subsection{Floquet engineering of the effective mass \label{subsec:Floquet-engineering-of_effective_mass}}

The term Floquet engineering describes the willful tuning of the electron
dispersion by a coherent, oscillating electromagnetic field. For the
massive Dirac Hamitlonian of Eq. (\ref{eq:general_H}) considered
in the letter, we find that the dispersion in the presence of an oscillating
field with frequency $\Omega_{F}$ and amplitude $\mathcal{E}$, is
given by
\begin{equation}
\varepsilon_{k}\approx\sqrt{\left(\left|\mathbf{d}\right|-\hbar\frac{\Omega_{F}}{2}\right)^{2}+\frac{e^{2}\mathcal{E}^{2}\lambda^{2}}{4\Omega_{F}^{2}\hbar^{2}}\left(2-\frac{E_{g}}{\left|\mathbf{d}\right|}\right)}.\label{eq:engineered_band}
\end{equation}
Here $\left|\mathbf{d}\right|=\sqrt{\lambda^{2}k^{2}+E_{g}^{2}/4}$.
We assumed circularly polarized light described by the vector potential
$\mathbf{A}=\left(\mathcal{E}/\Omega_{F}\right)\left[-\sin\Omega_{F}t,\cos\Omega_{F}t,0\right]$.
To derive Eq. (\ref{eq:engineered_band}) we changed to the rotating
frame by applying the unitary transformation $U=e^{i\hat{\mathbf{d}}\cdot\boldsymbol{\sigma}\Omega_{F}t}$
to Eq. (\ref{eq:general_H}) \citep{rudner2020band_engineering,lindner2011floquet}
and used the rotating wave approximation, ignoring all terms oscillating
at higher frequencies in the rotating frame \citep{rudner2020band_engineering,lindner2011floquet}.

To find the effective mass, we first consider the Fermi velocity,
which is given by
\begin{align*}
v_{F}\left(\mathcal{E}\right) & =\left.\frac{\partial\varepsilon_{k}}{\partial k}\right|_{k=k_{F}}\\
 & =\left.\frac{1}{\varepsilon_{k}}\left[\left(\left|\mathbf{d}\right|-\hbar\frac{\Omega_{F}}{2}\right)+\frac{e^{2}\mathcal{E}^{2}\lambda^{2}}{8\Omega_{F}^{2}\hbar^{2}}\frac{E_{g}}{\left|\mathbf{d}\right|^{2}}\right]\frac{\partial\left|\mathbf{d}\right|}{\partial k}\right|_{k=k_{F}}\\
 & =\frac{1}{\varepsilon_{k_{F}}}\left[\left(\left|\mathbf{d}\left(k_{F}\right)\right|-\hbar\frac{\Omega_{F}}{2}\right)+\frac{e^{2}\mathcal{E}^{2}\lambda^{2}}{8\Omega_{F}^{2}\hbar^{2}}\frac{E_{g}}{\left|\mathbf{d}\left(k_{F}\right)\right|^{2}}\right]\frac{\lambda^{2}k_{F}}{\left|\mathbf{d}\left(k_{F}\right)\right|}.
\end{align*}
For the effective mass, we find
\begin{align*}
m^{*}\left(\mathcal{E}\right) & =\frac{\hbar k_{F}}{v_{F}}=\frac{\hbar k_{F}}{\frac{1}{\varepsilon_{k_{F}}}\left[\left(\left|\mathbf{d}\left(k_{F}\right)\right|-\hbar\frac{\Omega_{F}}{2}\right)+\frac{e^{2}\mathcal{E}^{2}\lambda^{2}}{8\Omega_{F}^{2}\hbar^{2}}\frac{E_{g}}{\left|\mathbf{d}\left(k_{F}\right)\right|^{2}}\right]\frac{\lambda^{2}k_{F}}{\left|\mathbf{d}\left(k_{F}\right)\right|}}.
\end{align*}
For an amplitude modulated driving of the form $\mathcal{E}\left(t\right)=\bar{\mathcal{E}}+\delta\mathcal{E}\cos\left(2\omega_{1}t\right)$,
the parameter $h$ can be estimated as
\begin{equation}
h=\frac{1}{m^{*}\left(\bar{\mathcal{E}}\right)}\frac{\partial m^{*}\left(\bar{\mathcal{E}}\right)}{\partial\bar{\mathcal{E}}}\delta\mathcal{E}.\label{eq:h_eq}
\end{equation}
The behavior of the effective mass on the driving field strength $\mathcal{E}$
is thus a complicated function of the material and driving parameters.
In particular, it is non-analytic in $\mathcal{E}=0$ for $\left|\mathbf{d}\left(k_{F}\right)\right|=\hbar\frac{\Omega_{F}}{2}$,
where the electron quasi-dispersion exhibits a van-Hove singularity.
As a rule of thumb, we find from Eq. (\ref{eq:h_eq}) that away from
this singular point, the impact of the driving field on the effective
mass, and consequently, the effect of the modulation of $\mathcal{E}$
as parametrized by $h$, increases with a larger group velocity $v_{g}=\lambda/\hbar$,
a small gap size $E_{g}$, and a smaller driving frequency $\Omega_{F}$.

\subsection{Momentum current relationship in a Dirac system \label{subsec:Momentum-current-relationship}}

The relationship between current and momentum for a gapped Dirac Hamiltonian
can be established using the Boltzmann transport theory for Fermi
liquids. The Boltzmann equation for electrons reads \citep{abrikosov1959}
\begin{equation}
\partial_{t}f_{\mathbf{k},\lambda}+v_{\mathbf{k},\lambda}\cdot\nabla f_{\mathbf{k},\lambda}=\mathcal{C}_{\mathrm{coll}}\left[f\right]_{\mathbf{k},\lambda},
\end{equation}
with 
\begin{equation}
f_{\mathbf{k},\lambda}=\frac{1}{1+e^{\beta\left(\varepsilon_{\mathbf{k},\lambda}-\varepsilon_{F}-\hbar\mathbf{u}\cdot\mathbf{k}\right)}},
\end{equation}
where $\varepsilon_{\mathbf{k},\lambda}$ and $v_{\mathbf{k},\lambda}$
are the electron dispersion and group velocity, respectively, $\lambda$
is the band index, $\mathcal{C}_{\mathrm{coll}}$ is the collision
integral, and $\mathbf{u}$ is the flow velocity of the electrons.
In a metallic system as considered here, we can expand the dispersion
relation around the Fermi energy $\mathbf{\varepsilon}_{F}$, which
lies in the upper band ($\lambda=+$):
\begin{equation}
\varepsilon_{\mathbf{k},\lambda}\approx\varepsilon_{F}+\hbar v_{F}\left(k-k_{F}\right).\label{eq:energy_fermi_expand}
\end{equation}
To simplify the calculations, it is convenient to further approximate
Eq. (\ref{eq:energy_fermi_expand}) by a parabola, and write $\varepsilon_{\mathbf{k},\lambda}\approx\frac{\hbar^{2}k^{2}}{2m^{*}}\approx\varepsilon_{F}+\frac{\hbar^{2}k_{F}}{m^{*}}\left(k-k_{F}\right)$,
where
\[
m^{*}=\frac{\hbar k_{F}}{v_{F}}.
\]
We can now calculate the momentum density of the system
\begin{align}
\mathbf{p}\left(\mathbf{x},t\right) & =\sum_{\lambda}\int\frac{d^{2}k}{\left(2\pi\right)^{2}}\hbar\mathbf{k}f_{\mathbf{k},\lambda}\\
 & =\int\frac{d^{2}k}{\left(2\pi\right)^{2}}\hbar\mathbf{k}\frac{1}{1+e^{\beta\left(\varepsilon_{\mathbf{k},+}-\varepsilon_{F}-\hbar\mathbf{u}\left(\mathbf{x},t\right)\cdot\mathbf{k}\right)}}\\
 & \approx\int\frac{d^{2}k}{\left(2\pi\right)^{2}}\hbar\mathbf{k}\frac{1}{1+e^{\beta\left(\frac{\left(\hbar\mathbf{k}-m^{*}\mathbf{u}\left(\mathbf{x}\right)\right)^{2}}{2m^{*}}-\varepsilon_{F}-\frac{1}{2}m^{*}u^{2}\left(\mathbf{x},t\right)\right)}}\\
 & \approx\int\frac{d^{2}k}{\left(2\pi\right)^{2}}\left(\hbar\mathbf{k}+m^{*}\mathbf{u}\left(\mathbf{x},t\right)\right)\frac{1}{1+e^{\beta\left(\varepsilon_{\mathbf{k},+}-\varepsilon_{F}-\frac{1}{2}m^{*}u^{2}\left(\mathbf{x},t\right)\right)}}.
\end{align}
Since the distribution function does not depend on the angle of $\mathbf{k}$,
the first term of the integrand cancells out. This leaves us, to leading
order in the small $u/v_{F}$, with
\begin{equation}
\mathbf{p}\left(\mathbf{x},t\right)=\bar{\rho}m^{*}\mathbf{u}\left(\mathbf{x},t\right),\label{eq:mom_rel}
\end{equation}
where $\bar{\rho}$ is the total electron density
\begin{equation}
\bar{\rho}=\int\frac{d^{2}k}{\left(2\pi\right)^{2}}\frac{1}{1+e^{\beta\left(\varepsilon_{\mathbf{k},+}-\varepsilon_{F}\right)}}.
\end{equation}
A similar calculation can be carried out to find the current density.
To first order in $\mathbf{u}$, we find
\begin{align}
\mathbf{j}\left(\mathbf{x},t\right) & =e\int\frac{d^{2}k}{\left(2\pi\right)^{2}}\frac{1}{\hbar}\frac{\partial\varepsilon_{\mathbf{k},+}}{\partial\mathbf{k}}f_{\mathbf{k},\lambda}\nonumber \\
 & \approx e\int\frac{d^{2}k}{\left(2\pi\right)^{2}}\frac{\hbar\mathbf{k}}{m^{*}}f_{\mathbf{k},\lambda}\nonumber \\
 & \approx e\bar{\rho}\mathbf{u}\left(\mathbf{x},t\right).\label{eq:cur_rel}
\end{align}
From Eqs. (\ref{eq:mom_rel}) and (\ref{eq:cur_rel}), we find
\[
\mathbf{j\left(\mathbf{x},t\right)}=\frac{e}{m^{*}}\mathbf{p}\left(\mathbf{x},t\right),
\]
which is the expression used to derive Eq. (\ref{eq:dyn_real_space})
of the main text.

\subsection{Solving the time varying plasmon equation \label{subsec:slowly_varying_envelope}}

Here we show details of our solution to Eq. (\ref{eq:time_varying_medium_eq})
of the main text using the ansatz
\begin{equation}
\delta\rho=a\left(t\right)\cos\left(\omega_{1}t\right)+b\left(t\right)\sin\left(\omega_{1}t\right).\label{eq:slowly_varying_ansatz-1}
\end{equation}
Inserting Eq. (\ref{eq:slowly_varying_ansatz-1}) into Eq. (\ref{eq:time_varying_medium_eq})
results in
\begin{align*}
\partial_{t}\left(1+\frac{\delta m^{*}\left(t\right)}{m^{*}}\right)\partial_{t}\delta\rho & =-2\omega_{1}h\sin\left(2\omega_{1}t\right)\delta\dot{\rho}+\left(1-h\cos\left(2\omega_{1}t\right)\right)\delta\ddot{\rho}\\
 & =-2\omega_{1}h\sin\left(\omega_{1}t\right)\left[\dot{a}\cos\left(\omega_{1}t\right)+\dot{b}\sin\left(\omega_{1}t\right)\right.\\
 & \qquad\left.-a\omega_{1}\sin\left(\omega_{1}t\right)+b\omega_{1}\cos\left(\omega_{1}t\right)\right]\\
 & \qquad+\left(1-h\cos\left(2\omega_{1}t\right)\right)\left[\ddot{a}\cos\left(\omega_{1}t\right)+\ddot{b}\sin\left(\omega_{1}t\right)\right.\\
 & \qquad-2\dot{a}\omega_{1}\sin\left(\omega_{1}t\right)+2\dot{b}\omega_{1}\cos\left(\omega_{1}t\right)\\
 & \qquad\left.-a\omega_{1}^{2}\cos\left(\omega_{1}t\right)-b\omega_{1}^{2}\sin\left(\omega_{1}t\right)\right]
\end{align*}
\begin{align*}
\partial_{t}\left(1+\frac{\delta m^{*}\left(t\right)}{m^{*}}\right)\partial_{t}\delta\rho & \approx-\omega_{1}h\left[\dot{a}\sin\left(\omega_{1}t\right)+\dot{b}\cos\left(\omega_{1}t\right)\right.\\
 & \qquad\left.-a\omega_{1}\cos\left(\omega_{1}t\right)+b\omega_{1}\sin\left(\omega_{1}t\right)\right]\\
 & \qquad+\left[\ddot{a}\cos\left(\omega_{1}t\right)+\ddot{b}\sin\left(\omega_{1}t\right)\right.\\
 & \qquad-2\dot{a}\omega_{1}\sin\left(\omega_{1}t\right)+2\dot{b}\omega_{1}\cos\left(\omega_{1}t\right)\\
 & \qquad\left.-a\omega_{1}^{2}\cos\left(\omega_{1}t\right)-b\omega_{1}^{2}\sin\left(\omega_{1}t\right)\right]\\
 & \qquad+\left(h/2\right)\left[\ddot{a}\cos\left(\omega_{1}t\right)-\ddot{b}\sin\left(\omega_{1}t\right)\right.\\
 & \qquad+2\dot{a}\omega_{1}\sin\left(\omega_{1}t\right)+2\dot{b}\omega_{1}\cos\left(\omega_{1}t\right)\\
 & \qquad\left.-a\omega_{1}^{2}\cos\left(\omega_{1}t\right)+b\omega_{1}^{2}\sin\left(\omega_{1}t\right)\right]
\end{align*}
\begin{align*}
\gamma\partial_{t}\delta\rho & =\gamma\left[\dot{a}\cos\left(\omega_{1}t\right)+\dot{b}\sin\left(\omega_{1}t\right)\right.\\
 & \quad\left.-a\omega_{1}\sin\left(\omega_{1}t\right)+b\omega_{1}\cos\left(\omega_{1}t\right)\right].
\end{align*}
Comparing the coefficients in front of the sine and cosine functions
we find the equations for the amplitudes $a\left(t\right)$ and $b\left(t\right)$:
\begin{align*}
-\omega_{1}h\left(\dot{a}+\omega_{1}b\right)+\ddot{b}-2\dot{a}\omega_{1}-b\omega_{1}^{2}+\left(h/2\right)\left(-\ddot{b}+2\dot{a}\omega_{1}+b\omega_{1}^{2}\right)+\gamma\left(\dot{b}-a\omega_{1}\right)+\omega_{\mathrm{pl}}^{2}\left(q\right)b & =0\\
-\omega_{1}h\left(\dot{b}-\omega_{1}a\right)+\ddot{a}+2\dot{b}\omega_{1}-a\omega_{1}^{2}+\left(h/2\right)\left(\ddot{a}+2\dot{b}\omega_{1}-a\omega_{1}^{2}\right)+\gamma\left(\dot{a}+b\omega_{1}\right)+\omega_{\mathrm{pl}}^{2}\left(q\right)a & =0
\end{align*}
anticipating $a\sim b\sim e^{st}$, where $s\sim\omega_{1}h$ and
ignoring $\mathcal{O}\left(h^{2}\right)$ and higher, we find
\begin{align}
-\omega_{1}^{2}hb/2-2\dot{a}\omega_{1}-\gamma a\omega_{1}+\left(\omega_{\mathrm{pl}}^{2}\left(q\right)-\omega_{1}^{2}\right)b & =0\nonumber \\
\omega_{1}^{2}ha/2+2\dot{b}\omega_{1}+\gamma b\omega_{1}+\left(\omega_{\mathrm{pl}}^{2}\left(q\right)-\omega_{1}^{2}\right)a & =0.\label{eq:a_b_lin_eom-1}
\end{align}
Next we use the ansatz $a=\tilde{a}e^{st}$, $b=\tilde{b}e^{st}$:
\[
\left[\begin{array}{cc}
-2s\omega_{1}-\gamma\omega_{1} & -\frac{1}{2}\omega_{1}^{2}h+\left(\omega_{\mathrm{pl}}^{2}-\omega_{1}^{2}\right)\\
\frac{1}{2}\omega_{1}^{2}h+\left(\omega_{\mathrm{pl}}^{2}-\omega_{1}^{2}\right) & 2s\omega_{1}+\gamma\omega_{1}
\end{array}\right]\left[\begin{array}{c}
a\\
b
\end{array}\right]=0.
\]
The solvability condition
\[
-\left(2s+\gamma\right)^{2}-\frac{\left(\omega_{\mathrm{pl}}^{2}-\omega_{1}^{2}\right)^{2}}{\omega_{1}^{2}}+\left(\frac{1}{2}\omega_{1}h\right)^{2}=0
\]
gives
\[
s=-\frac{\gamma}{2}\pm\frac{1}{2}\sqrt{\left(\frac{1}{2}\omega_{1}h\right)^{2}-\frac{\left(\omega_{\mathrm{pl}}^{2}-\omega_{1}^{2}\right)^{2}}{\omega_{1}^{2}}}.
\]
This corresponds to Eq. (\ref{eq:gap_bands}) of the main text. The
condition of instability is $s>0$:
\[
\gamma<\sqrt{\left(\frac{1}{2}\omega_{1}h\right)^{2}-\frac{\epsilon^{4}}{\omega_{1}^{2}}},
\]
thus plasmons in the frequency region 
\[
\epsilon^{4}<\frac{1}{4}\omega_{1}^{4}h^{2}-\omega_{1}^{2}\gamma^{2}
\]
will become unstable if 
\[
h>\frac{2\gamma}{\omega_{1}}.
\]
The instability grows according to
\begin{equation}
\text{\ensuremath{\delta\rho}}\sim e^{\left(h\omega_{1}/4-\gamma/2\right)t}.\label{eq:exp_plasmon_growth}
\end{equation}

\subsection{Production of entangled plasmon pairs \label{subsec:Production-of-entangled}}

\begin{figure}
\centering{}\includegraphics[scale=0.45]{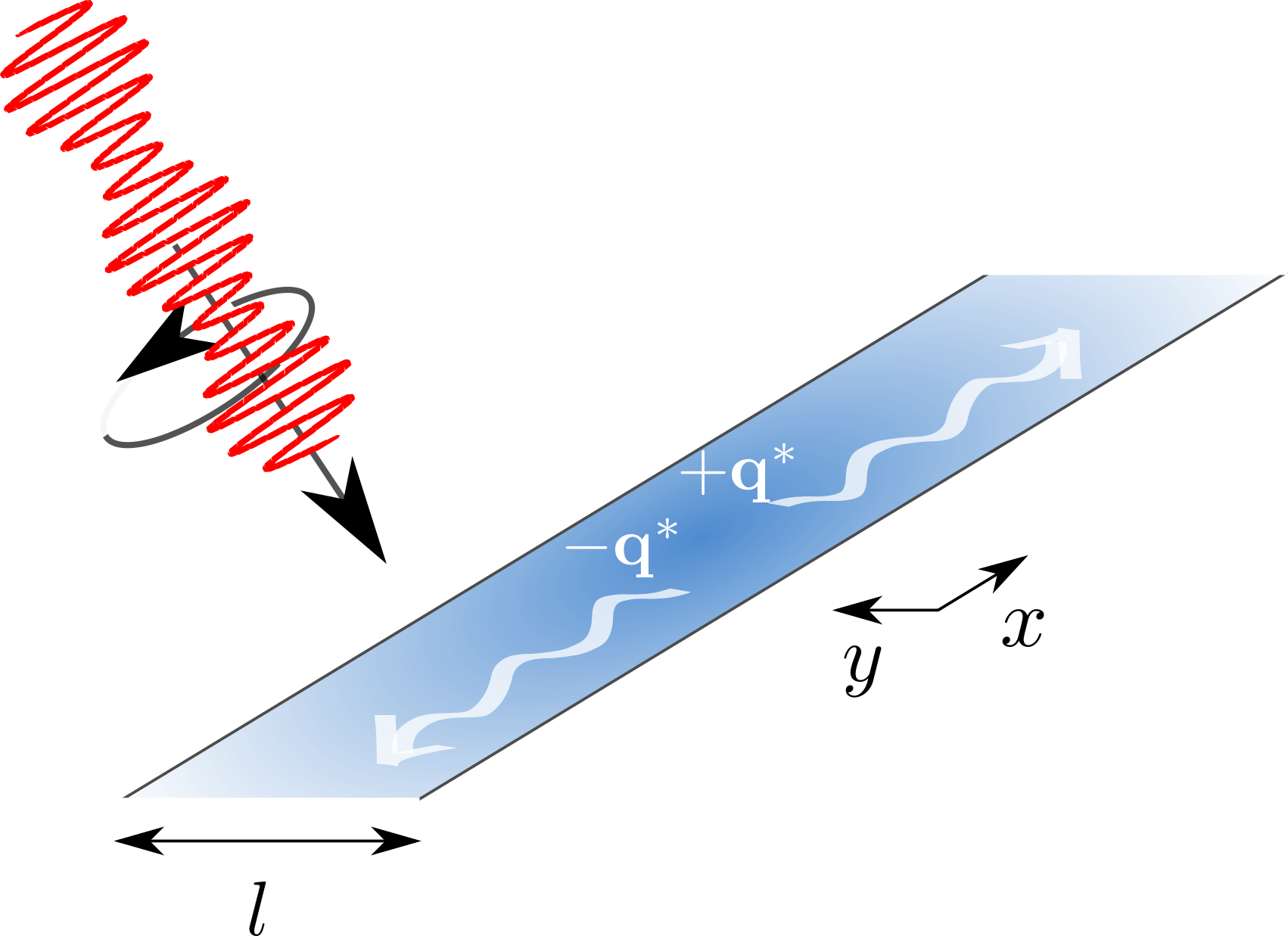}\caption{Quasi one-dimensional strip geometry considered in Eq. (\ref{eq:quasi-1d-pot}).
Due to momentum conservation, plasmons created by MFPD come in entangled
pairs with momenta $\pm q^{*}$. \label{fig:slab}}
\end{figure}
Eq (\ref{eq:time_varying_medium_eq}) of the main text -- the equation
of motion for the charge density -- can be derived from the Hamiltonian
\begin{equation}
H_{\mathrm{pl}}=\sum_{\mathbf{q}}\frac{1}{2}\left(1+h\cos(2\omega_{1}t)\right)^{-1}p_{\mathbf{q}}p_{-\mathbf{q}}+\frac{1}{2}\omega_{\mathrm{pl}}^{2}\left(q\right)\delta\rho_{\mathbf{q}}\delta\rho_{-\mathbf{q}}\label{eq:Entangled_Ham_Suppl}
\end{equation}
using Hamilton's equations:
\begin{equation}
\begin{array}{c}
\partial_{t}\delta\rho_{\mathbf{q}}=\frac{\partial H}{\partial p_{\mathbf{q}}}=\frac{p_{-\mathbf{q}}}{1+h\cos(2\omega_{1}t)}\\
\partial_{t}p_{\mathbf{q}}=-\frac{\partial H}{\partial\delta\rho_{\mathbf{q}}}=-\omega_{\mathrm{pl}}^{2}\left(q\right)\delta\rho_{-\mathbf{q}}
\end{array}.
\end{equation}
In the following, we focus on the quasi-1D geometry of Fig.\ref{fig:slab}
where the wavevectors $\mathbf{q}$ are replaced by the wavenumbers
$\pm q$. We subject the Hamiltonian (\ref{eq:Entangled_Ham_Suppl})
to canonical quantization. Creation and anihillation operators are
defined as

\begin{equation}
\begin{array}{c}
a_{q}=\sqrt{\frac{\omega_{\mathrm{pl}}\left(q\right)}{2\hbar}}\left(\delta\rho_{q}-\frac{p_{-q}}{i\omega_{\mathrm{pl}}\left(q\right)}\right)\\
a_{q}^{\dagger}=\sqrt{\frac{\omega_{\mathrm{pl}}\left(q\right)}{2\hbar}}\left(\delta\rho_{-q}+\frac{p_{q}}{i\omega_{\mathrm{pl}}\left(q\right)}\right),
\end{array}
\end{equation}
and canonical quantization is implemented by demanding 
\begin{equation}
\left[a_{q},a_{q'}^{\dagger}\right]=\delta_{q,q'}.
\end{equation}
Written in terms of creation and annihilation operators, the Hamiltonian
of Eq. (\ref{eq:Entangled_Ham_Suppl}) reads
\begin{equation}
\begin{array}{c}
H=\sum_{q}\frac{\hbar\omega_{pl}(q)}{4}\left(1+\frac{1}{1+h\cos(2\omega_{1}t)}\right)\left(a_{q}^{\dagger}a_{q}+a_{-q}^{\dagger}a_{-q}\right)\\
+\frac{\hbar\omega_{pl}(q)}{4}\left(\frac{h\cos(2\omega_{1}t)}{1+h\cos(2\omega_{1}t)}\right)\left(a_{q}^{\dagger}a_{-q}^{\dagger}+a_{-q}a_{q}\right).
\end{array}
\end{equation}
Focusing on the resonant mode with $q=q^{*}$, for a small modulation
strength $h\ll1$, we find
\begin{equation}
H_{\mathrm{pl},\mathrm{int}}\approx\frac{h}{4}\hbar\omega_{1}\left(a_{q^{*}}^{\dagger}a_{-q^{*}}^{\dagger}+a_{-q^{*}}a_{q^{*}}\right)\label{eq:Interaction_h_gen_plasmons-1}
\end{equation}
for the Hamiltonian in the interaction representation.

Applying the time evolution operator $U_{\mathrm{pl,int}}\left(t,0\right)=\exp\left(-tH_{\mathrm{pl},\mathrm{int}}/\hbar\right)$
to the vacuum generates a non-factorizable two mode squeezed state
$\ket{\psi_{\mathrm{pl}}\left(t\right)}=U_{\mathrm{pl,int}}\left(t,0\right)\ket{0,0}$
in the basis $\ket{n_{q},n_{-q}}$ \citep{gerry2005introductory_quantum_optics}.
Here, $n_{q}$ is the number of plasmons in the state $q$:
\begin{equation}
\ket{\psi_{\mathrm{pl}}\left(t\right)}=\frac{1}{\cosh\left(\frac{h\omega_{1}t}{4}\right)}\sum_{n=0}\tanh^{n}\left(\frac{h\omega_{1}t}{4}\right)|n_{q},n_{-q}\rangle.
\end{equation}
The total number of plasmons is given by
\begin{equation}
\langle N_{\pm q}\rangle=\bra{\psi_{\mathrm{pl}}\left(t\right)}a_{\pm q}^{\dagger}a_{\pm q}\ket{\psi_{\mathrm{pl}}\left(t\right)}.
\end{equation}
We find
\begin{equation}
\langle N_{\pm q}\rangle=\sinh^{2}\left(\frac{h\omega_{1}t}{4}\right)\sim\exp\left(\frac{h\omega_{1}t}{2}\right),
\end{equation}
which means that the intensity of the plasmon fields grows in accordance
with the classical result of Eq. (\ref{eq:gap_bands}) of the main
text.

\subsection{Pulse induced time reversal \label{subsec:Pulse-induced-time_reversal}}

Let us consider an optical pulse with frequency $\Omega_{F}$ and
envelope $\mathcal{E}\left(t\right)=\mathcal{E}_{0}f\left(t-t_{0}\right)$,
such that the maximum amplitude $\mathcal{E}_{0}$ is reached at $t_{0}$.
Let the pulse have a width $\Delta t$, which is chosen such that
$2\pi/\Omega_{F}\ll\Delta t\ll2\pi/\omega_{\mathrm{pl}}\left(q_{0}\right)$,
where $\omega_{\mathrm{pl}}\left(q_{0}\right)$ is the central frequency
of the plasmon wave packet. The dependence of the effective mass on
$\mathcal{E}$ is given by the formula 
\begin{equation}
m^{*}\left(t\right)=\left.\frac{\varepsilon_{k}\left(\mathcal{E}\left(t\right),\Omega_{F}\right)}{\lambda k_{F}^{2}\left(1-\frac{\hbar\Omega_{F}}{2\left|\mathbf{d}\right|}\right)}\right|_{k=k_{F}}.
\end{equation}
For simplicity, we can assume that 
\begin{equation}
f\left(t\right)=\Theta\left(t-\Delta t/2\right)\Theta\left(t+\Delta t/2\right),
\end{equation}
which translates to 
\begin{equation}
m^{*}\left(t\right)=\bar{m}^{*}+f\left(t\right)\left.\frac{\varepsilon_{k}\left(\mathcal{E}_{0},\Omega_{F}\right)}{\lambda k_{F}^{2}\left(1-\frac{\hbar\Omega_{F}}{2\left|\mathbf{d}\right|}\right)}\right|_{k=k_{F}}.
\end{equation}
If the pulse is very short with $\Delta t\ll2\pi/\omega_{\mathrm{pl}}\left(q_{0}\right)$,
we can further approximate it with a delta function: $f\left(t\right)\approx\Delta t\delta\left(t-t_{0}\right)$.
This yields
\begin{equation}
m^{*}\left(t\right)=\bar{m}^{*}+\Delta t\delta\left(t-t_{0}\right)\left.\frac{\varepsilon_{k}\left(\mathcal{E}_{0},\Omega_{F}\right)}{\lambda k_{F}^{2}\left(1-\frac{\hbar\Omega_{F}}{2\left|\mathbf{d}\right|}\right)}\right|_{k=k_{F}}.
\end{equation}
Remembering that, although the pulse is short at the plasmon time-scale,
it is actually long compared to $2\pi/\Omega_{F}$, we can write
\begin{equation}
m^{*}\left(t\right)=\bar{m}^{*}\left(1+\bar{h}\delta\left(t-t_{0}\right)\right),
\end{equation}
where
\begin{align}
\bar{h} & =\left.\frac{\Delta t\varepsilon_{k}\left(\mathcal{E}_{0},\Omega_{F}\right)}{\lambda\bar{m}^{*}k_{F}^{2}\left(1-\frac{\hbar\Omega_{F}}{2\left|\mathbf{d}\right|}\right)}\right|_{k=k_{F}}
\end{align}
is an amplitude characterizing the pulse. Thus, for a single driving
pulse, Eq. (\ref{eq:time_varying_medium_eq}) becomes
\begin{equation}
\partial_{t}^{2}\delta\rho_{\mathbf{q}}+\omega_{\mathrm{pl}}^{2}\left(q\right)\delta\rho{}_{\mathbf{q}}=-\bar{h}\partial_{t}\delta\left(t-t_{0}\right)\partial_{t}\delta\rho_{\mathbf{q}}.\label{eq:wave_with_inverting_inhomog-1}
\end{equation}
For $t<t_{0}$, we consider a propagating wave-packet of the form
\begin{equation}
\delta\rho_{i}\left(\mathbf{x},t\right)=\int\frac{d^{2}q}{2\pi}\phi_{i}\left(\mathbf{q}\right)e^{i\mathbf{q}\cdot\mathbf{x}-i\omega_{\mathrm{pl}}\left(q\right)t},
\end{equation}
which is a linear superposition of solutions to the homogeneous part
of Eq. (\ref{eq:wave_with_inverting_inhomog-1}). For $t\geq t_{0}$,
we write
\begin{equation}
\delta\rho\left(\mathbf{x},t\right)=\int\frac{d^{2}q}{2\pi}\phi\left(\mathbf{q}\right)e^{i\mathbf{q}\cdot\mathbf{x}-i\omega_{\mathrm{pl}}\left(q\right)t}
\end{equation}
To find the solution for $t>t_{0}$, we use the retarded Green's function
\begin{equation}
G_{R}\left(t\right)=\Theta\left(t\right)\frac{\sin\left(\omega\left(q\right)t\right)}{\omega\left(q\right)},
\end{equation}
which fulfills 
\begin{equation}
\partial_{t}^{2}G_{R}\left(t\right)+\omega_{\mathrm{pl}}^{2}\left(q\right)G_{R}\left(t\right)=\delta\left(t\right).
\end{equation}
From this expression, it is evident that a solution to Eq. (\ref{eq:wave_with_inverting_inhomog-1})
is given by
\begin{equation}
-i\int\frac{d^{2}q}{2\pi}\int_{-\infty}^{\infty}dt'\left[\partial_{t'}G_{R}\left(t-t'\right)\right]\bar{h}\delta\left(t'-t_{0}\right)\phi\left(\mathbf{q}\right)\omega_{\mathrm{pl}}\left(q\right)e^{i\mathbf{q}\cdot\mathbf{x}-i\omega_{\mathrm{pl}}\left(q\right)t'}
\end{equation}
where we used integration by parts. In the above equation, the term
involving the derivative of the Heaviside function inside $G_{R}$
vanishes, and we are left with
\begin{align}
 & i\int\frac{d^{2}q}{2\pi}\Theta\left(t-t_{0}\right)\cos\left(\omega\left(q\right)\left(t-t_{0}\right)\right)\bar{h}\omega_{\mathrm{pl}}\left(q\right)\phi\left(\mathbf{q}\right)e^{i\mathbf{q}\cdot\mathbf{x}-i\omega_{\mathrm{pl}}\left(q\right)t_{0}}\nonumber \\
 & \quad=\frac{i}{2}\bar{h}\Theta\left(t-t_{0}\right)\partial_{t}\left[\delta\rho\left(\mathbf{x},2t_{0}-t\right)+\delta\rho\left(\mathbf{x},t\right)\right]
\end{align}
The full solution is the sum of the homogeneous and pulse induced
parts:
\begin{equation}
\delta\rho\left(\mathbf{x},t\right)=\delta\rho_{i}\left(\mathbf{x},t\right)+\frac{i}{2}\bar{h}\Theta\left(t-t_{0}\right)\partial_{t}\left[\delta\rho\left(\mathbf{x},2t_{0}-t\right)+\delta\rho\left(\mathbf{x},t\right)\right].\label{eq:self_consistent_pulse_eq-1}
\end{equation}
Eq. (\ref{eq:self_consistent_pulse_eq-1}) is a self consistent equation
and can be solved by iteration. To first order in $\bar{h}$, we find
\begin{align}
\delta\rho\left(\mathbf{x},t\right) & \approx\delta\rho_{i}\left(\mathbf{x},t\right)+\frac{i}{2}\bar{h}\Theta\left(t-t_{0}\right)\partial_{t}\left[\delta\rho_{i}\left(\mathbf{x},2t_{0}-t\right)+\delta\rho_{i}\left(\mathbf{x},t\right)\right]
\end{align}
The time inverted, backwards propagating wave is given by
\begin{equation}
\delta\rho_{\mathrm{rev}}\left(\mathbf{x},t\right)\approx\frac{i}{2}\bar{h}\Theta\left(t-t_{0}\right)\partial_{t}\delta\rho_{i}\left(\mathbf{x},2t_{0}-t\right).
\end{equation}
For a narrow packet, where $\phi\left(\mathbf{q}\right)$ is centered
around a wavenumber $q_{0}$, and has a width $\Delta q$, expanding
to leading order in $\Delta q/q_{0}$, we have 
\begin{equation}
\delta\rho_{\mathrm{rev}}\left(\mathbf{x},t\right)\approx A\left(q_{0}\right)\delta\rho_{i}\left(\mathbf{x},-t\right)
\end{equation}
with a complex time-inversion amplitude 
\begin{equation}
A\left(q_{0}\right)=\frac{1}{2}\omega_{\mathrm{pl}}\left(q_{0}\right)\bar{h}e^{-i2\omega_{\mathrm{pl}}\left(q_{0}\right)t_{0}}.
\end{equation}

\begin{figure}
\centering{}\includegraphics[width=0.35\textwidth]{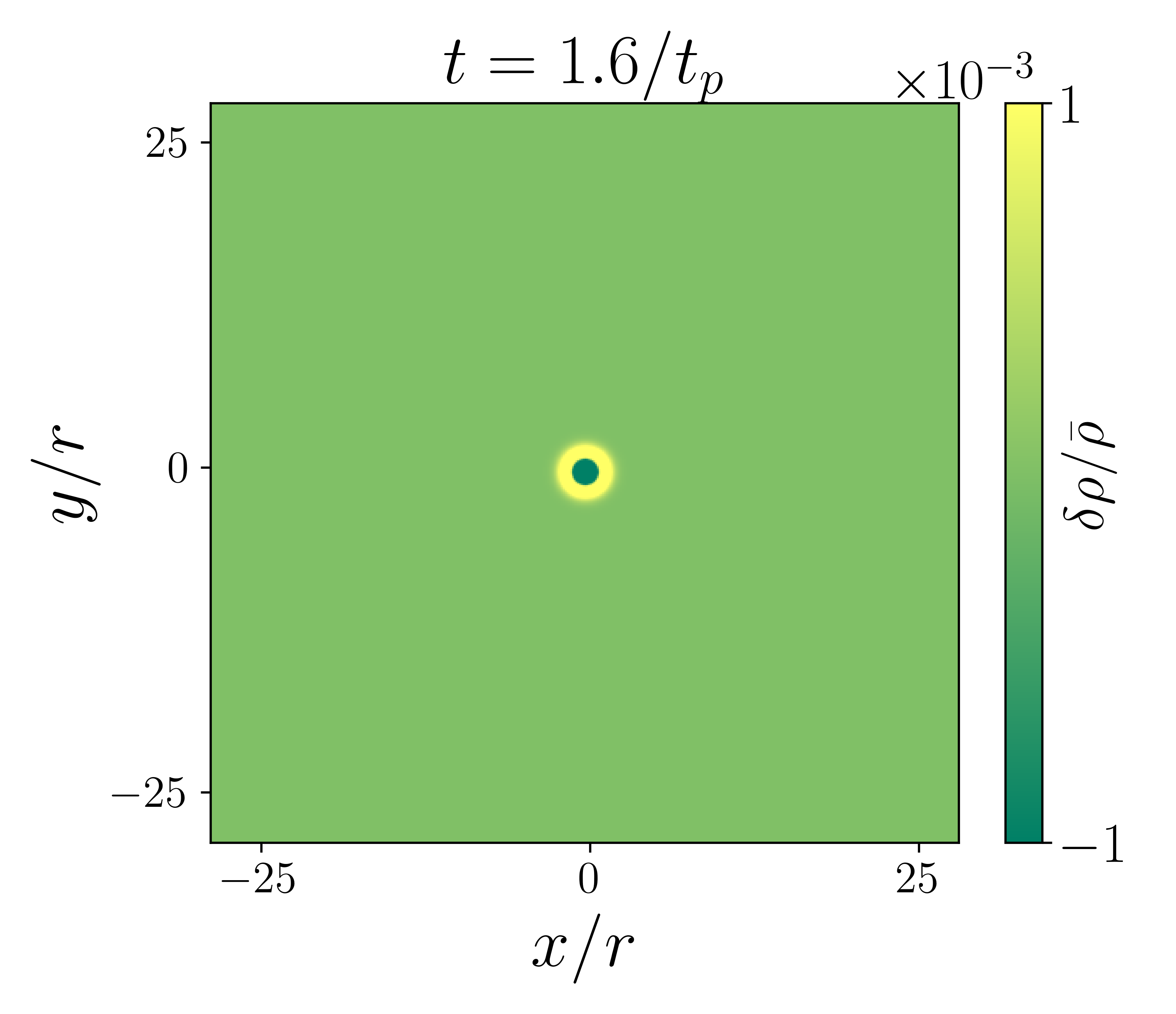}\includegraphics[width=0.35\textwidth]{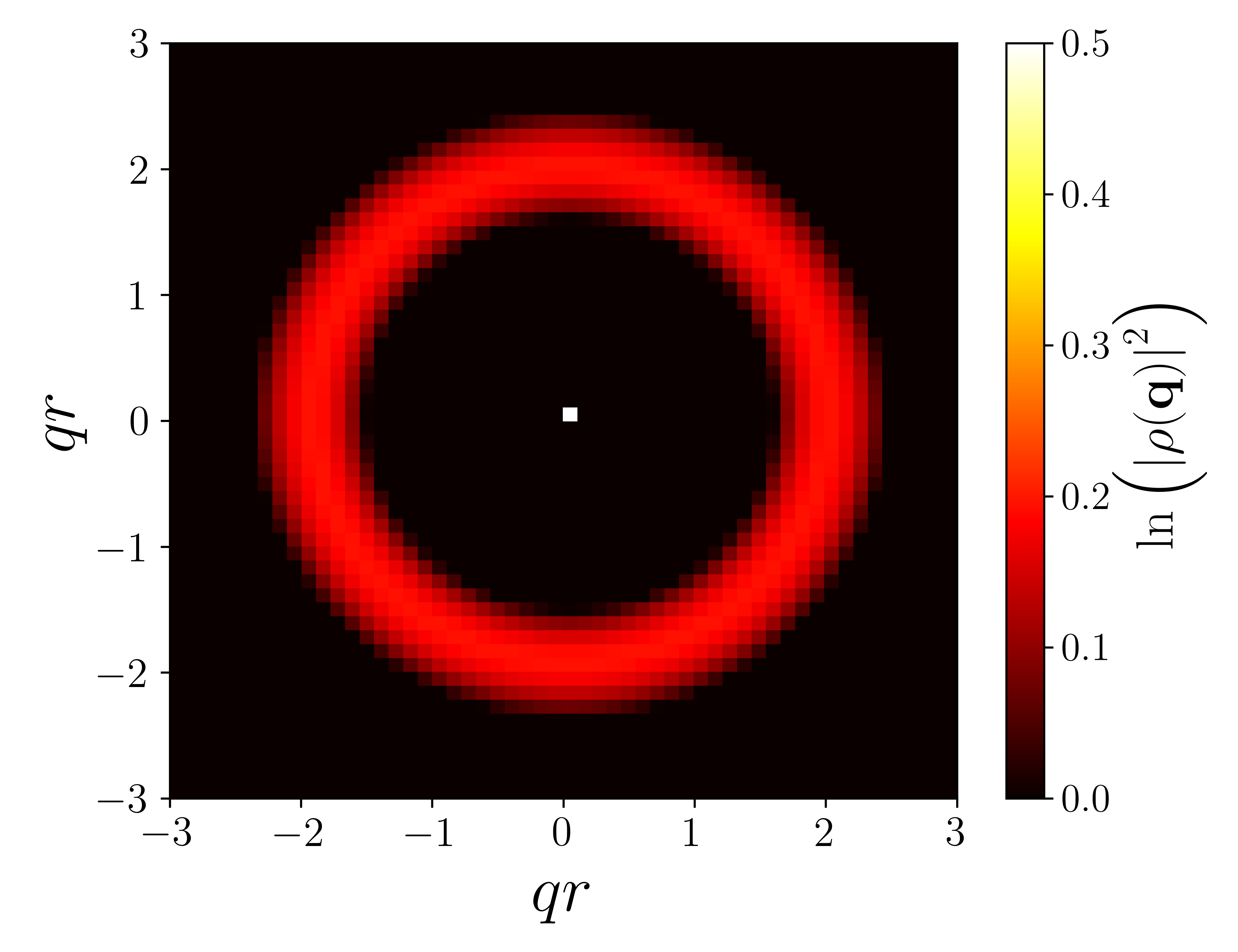}\caption{Plasmon wave at $t=3t_{p}$ and its Fourier transform.\label{fig:plasmon_fourier}}
\end{figure}

To explore the time-reversal effect in more detail and to consider
pulse durations which are more realistic than the infinitely short
pulse assumed above, we solve Eq. (\ref{eq:dyn_real_space}) numerically.
To create a cylindrical we use a potential of the form
\[
\phi\left(\mathbf{x},t\right)=Ce^{-\frac{x^{2}+y^{2}}{r^{2}}}te^{-t/t_{p}}.
\]
Here $r$ can be thought of as the radius of a tip used to force the
system and $t_{p}$ measures the duration of the pulse. At time $t_{0}=75t_{p}$,
we apply a strong sudden change of the effective mass:
\[
m^{*}\left(t\right)=\bar{m}^{*}\left(1+\bar{h}e^{-t^{2}/\Delta t^{2}}\right).
\]
We explore the effect of different pulse durations $\Delta t$ to
the time-reversed signal. It is important to compare $\Delta t$ to
the characteristic frequency range of the plasmon signal to which
pulse-induced time-reversal is applied. To this end, we perform a
Fourier transform of the signal at $t=3t_{p}$. The result is shown
in Fig. \ref{fig:plasmon_fourier}. The plasmon wavenumbers are roughly
peaked around $qr=1.9$. This means that the plasmon wavelength is
roughly set by the diameter of the tip. Using the dispersion relation
of Eq. (\ref{eq:equilib_plasmon_modes}), we find that a wavenumber
of $qr=1.9$ corresponds to an oscillation cycle of $T_{\mathrm{pl}}\left(q\right)=2\pi/\omega_{\mathrm{pl}}\left(q\right)\approx10t_{p}$.

\begin{figure}
\begin{centering}
\includegraphics[width=0.35\textwidth]{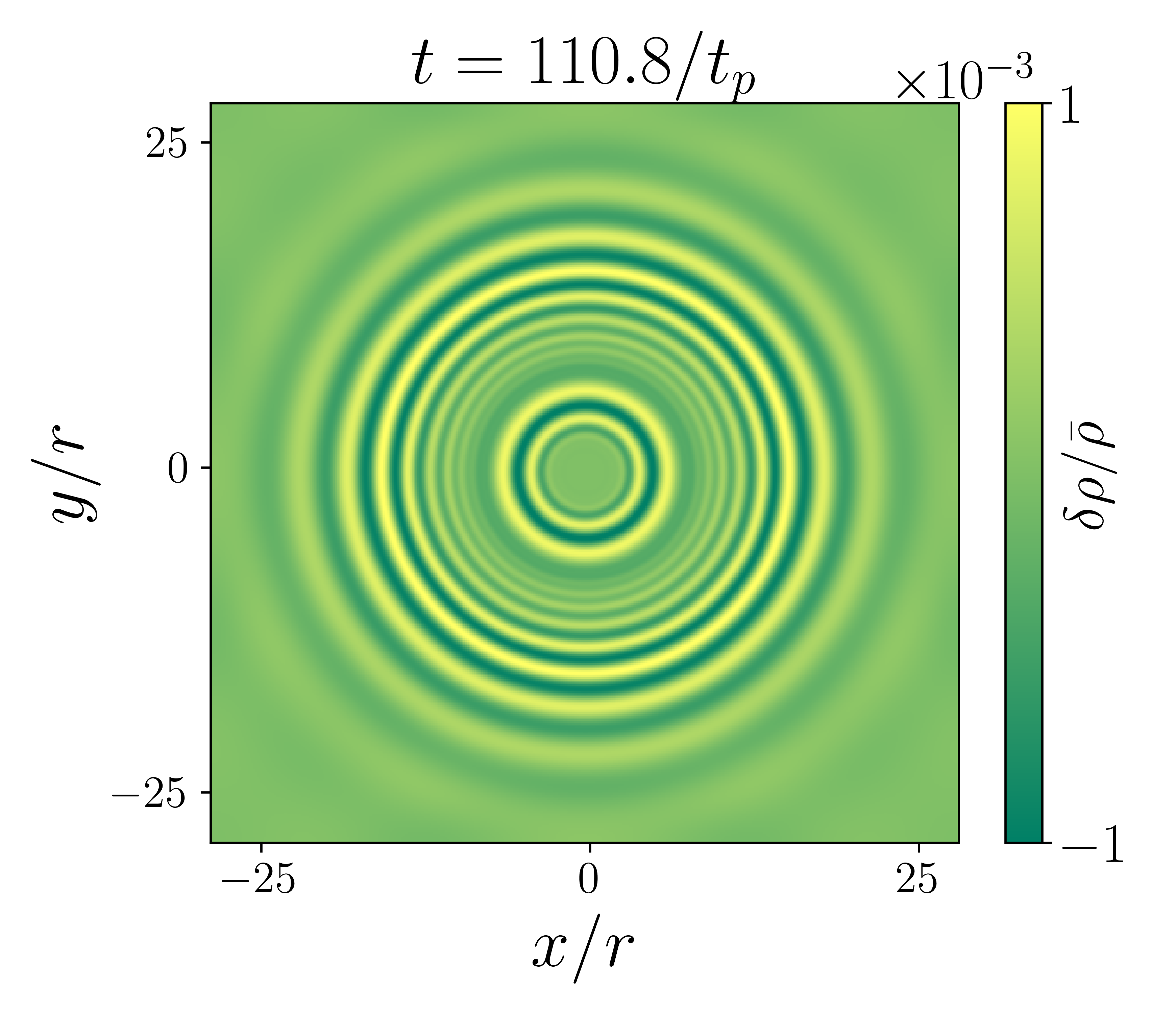}\includegraphics[width=0.35\textwidth]{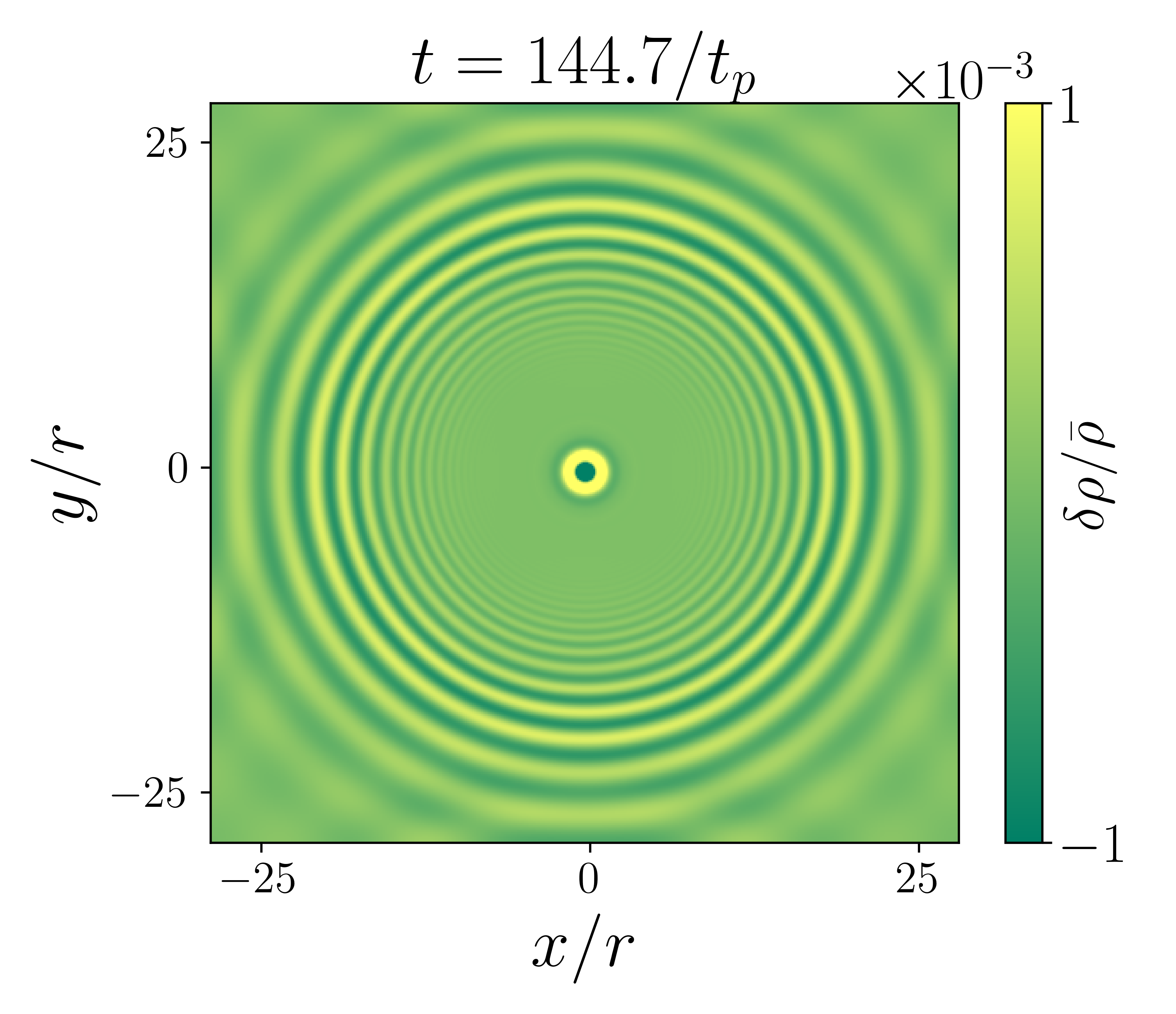}
\par\end{centering}
\begin{centering}
\includegraphics[width=0.35\textwidth]{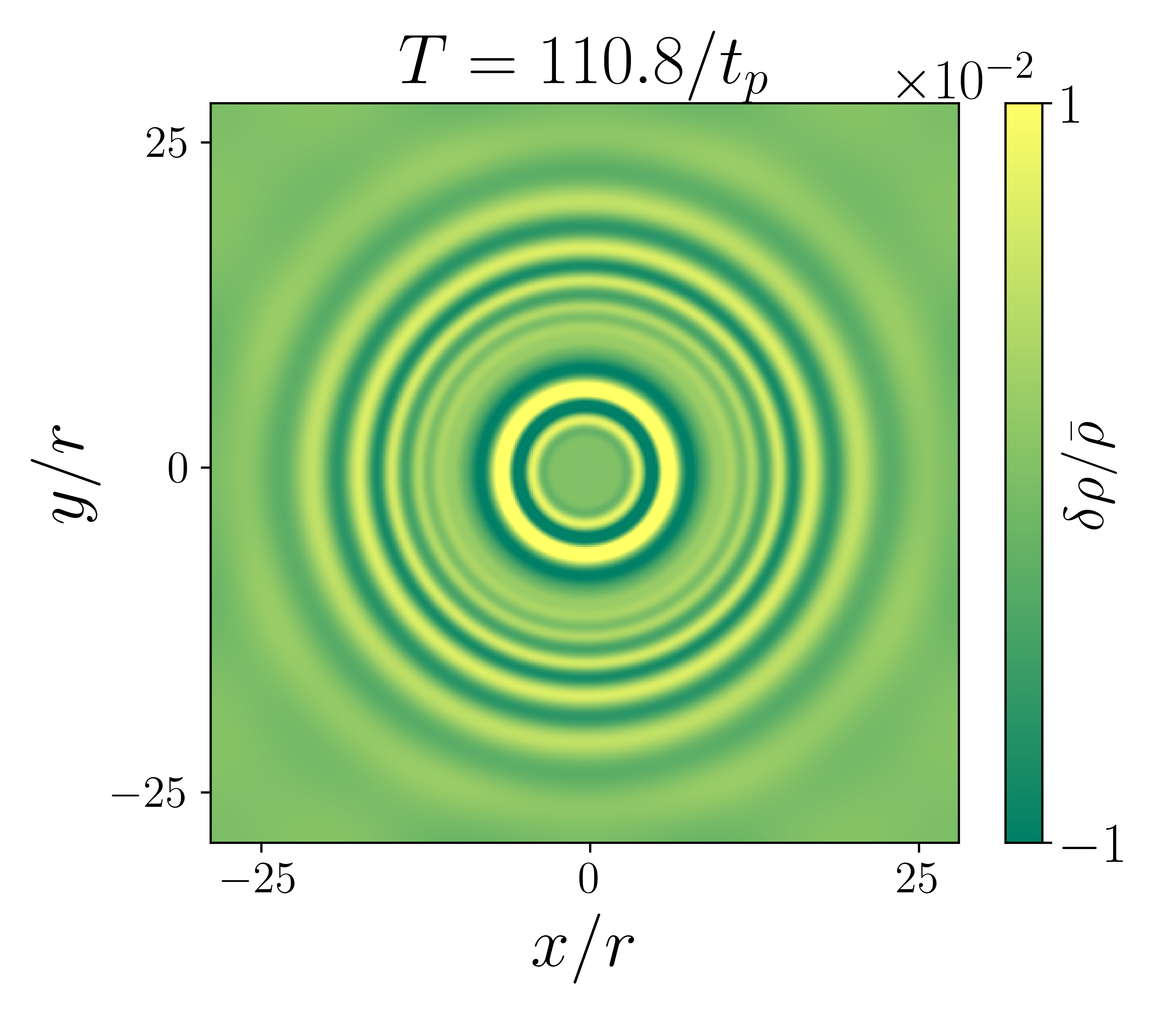}\includegraphics[width=0.35\textwidth]{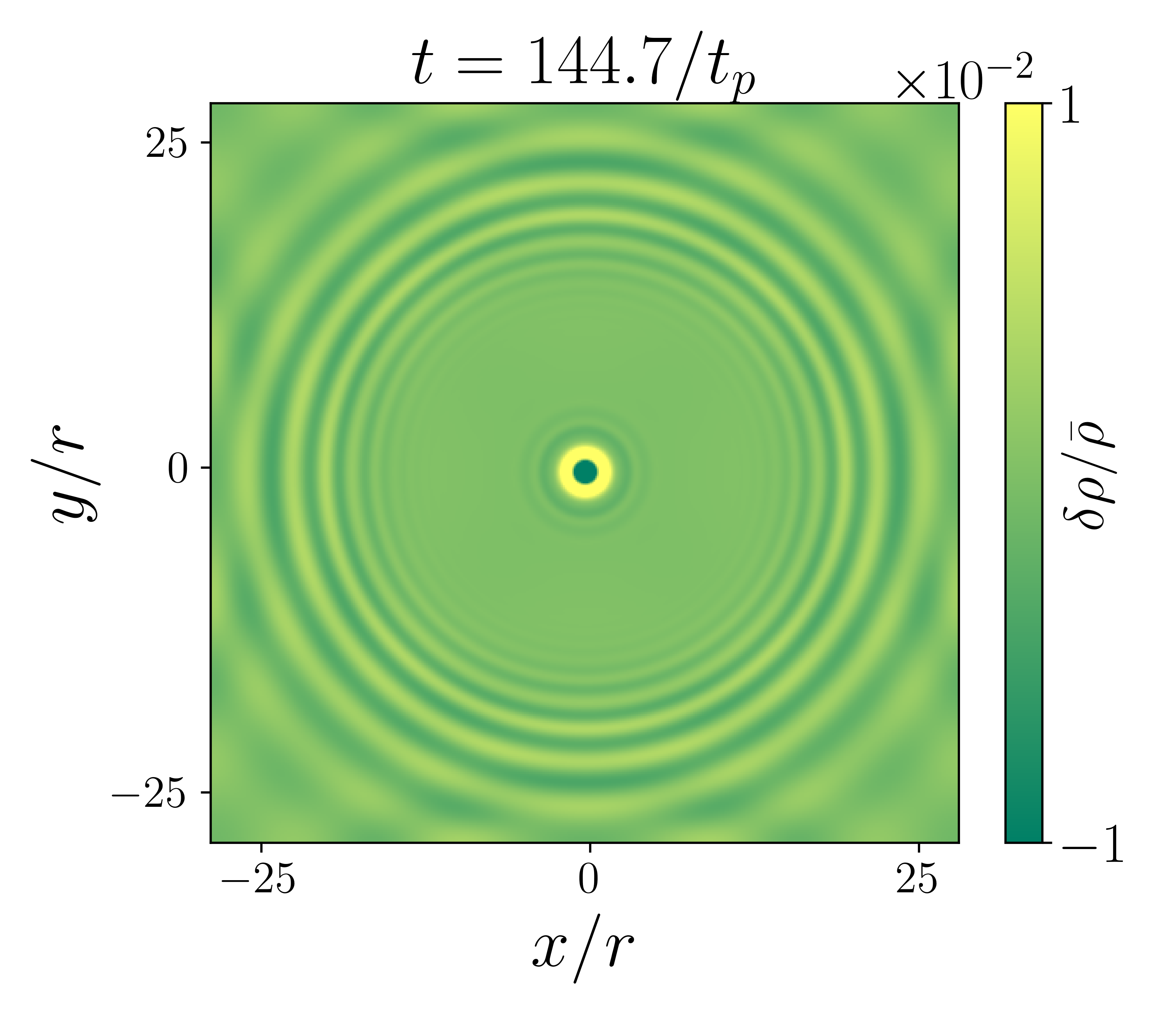}
\par\end{centering}
\centering{}\includegraphics[width=0.35\textwidth]{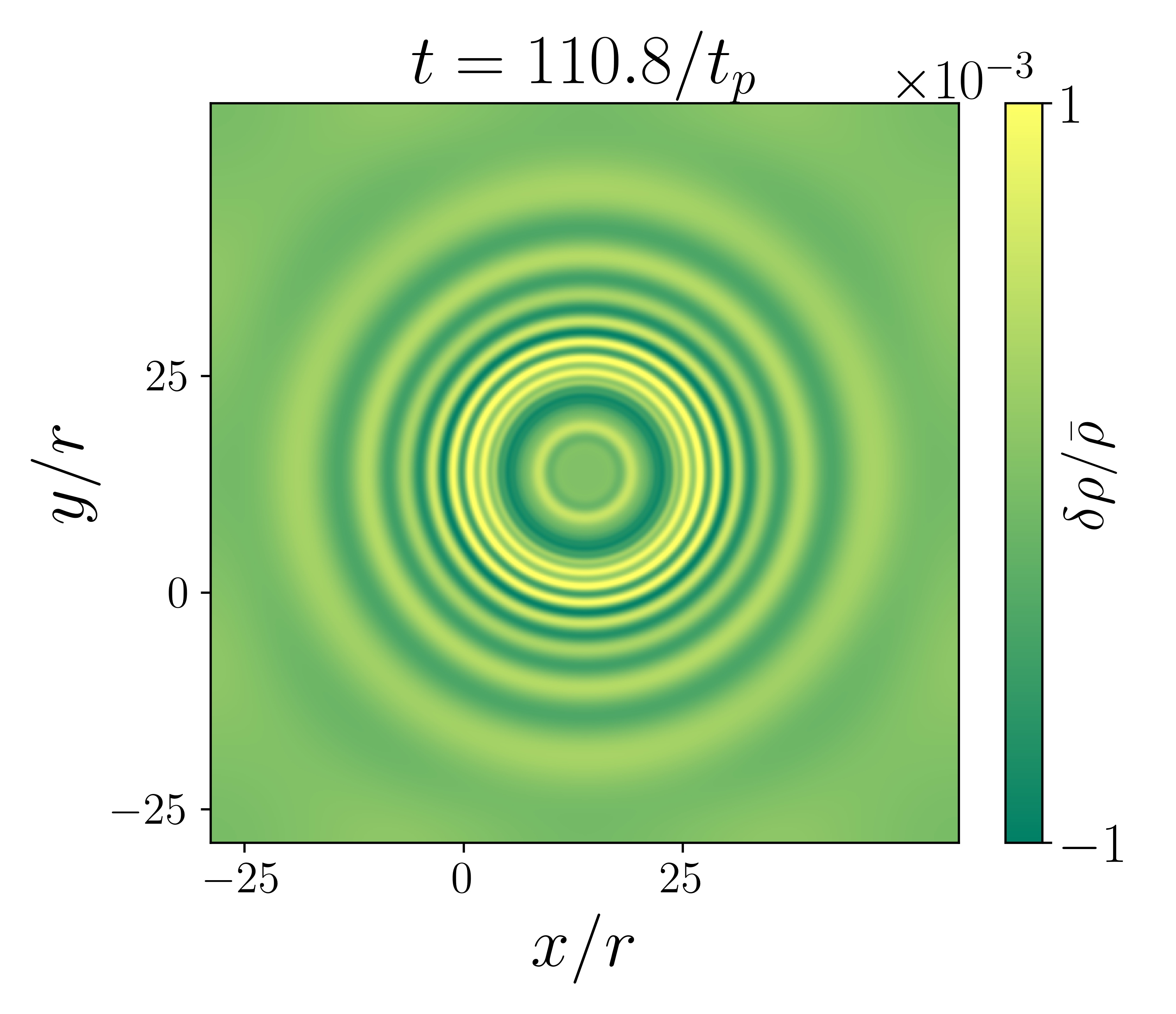}\includegraphics[width=0.35\textwidth]{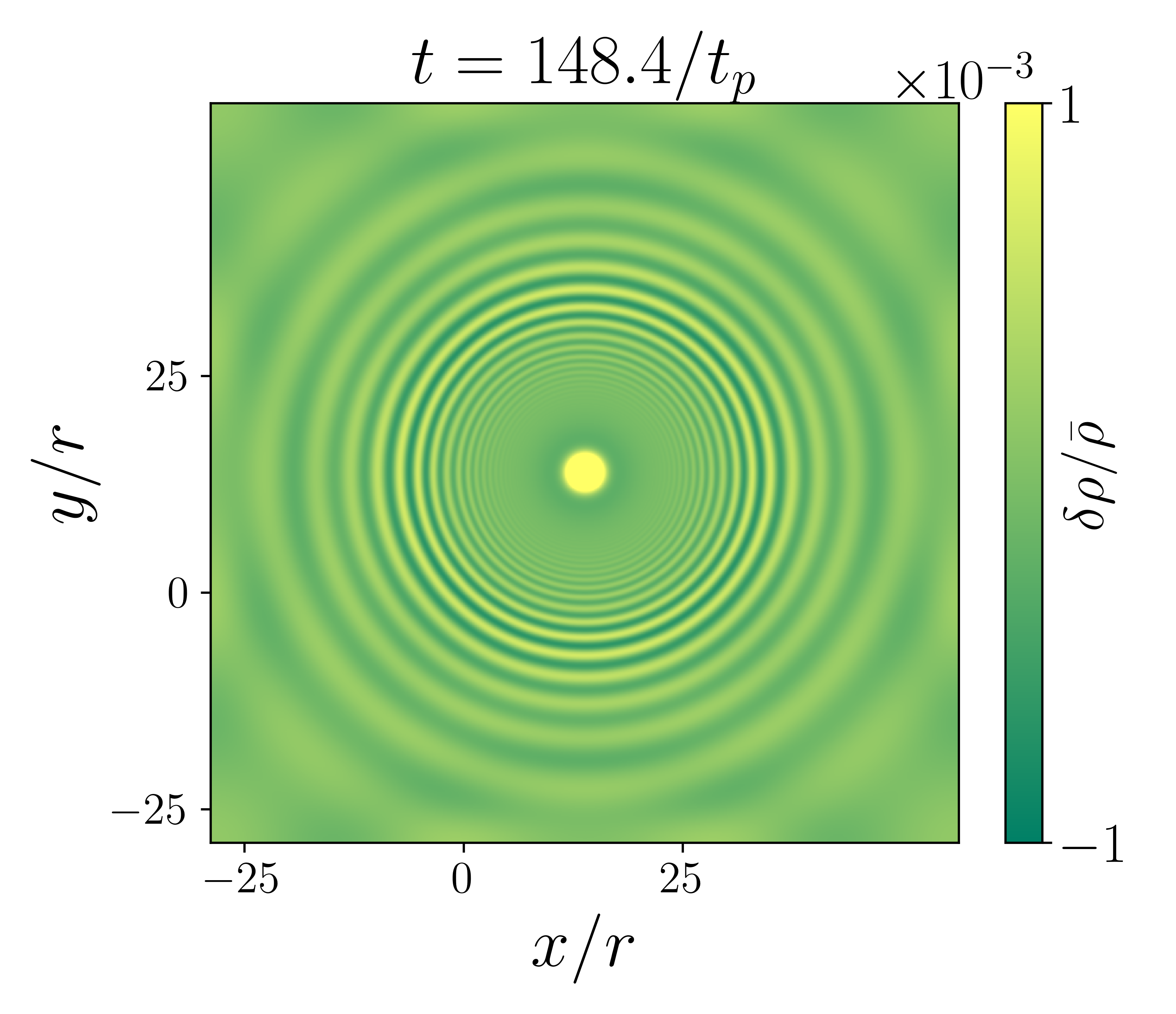}\caption{Pule induced time reversal with a pulse duration of $\Delta t=0.1T_{\mathrm{pl}}$
(first row), $\Delta t=0.22T_{\mathrm{pl}}$ (second row) and $\Delta t=0.5T_{\mathrm{pl}}$
(third row). The pulse was triggered at $t=75t_{p}.$ The inner concentric
circles represent the time-reversed wave. For $\Delta t=0.5T_{\mathrm{pl}}$,
we observe a strong interference of the forward and backward propagating
waves, however, a time-reversed signal is clearly observable at $t\approx150t_{p}$.
\label{fig:time_reversal_vs_pulse_duration}}
\end{figure}

We run simulations with $\Delta t=0.1T_{\mathrm{pl}}$ and $\Delta t=0.22T_{\mathrm{pl}}$
with $\bar{h}=1$ and $\Delta t=0.5T_{\mathrm{pl}}$ with $\bar{h}=0.2$.
Our main goal was to establish wether a pulse whose duration is not
by orders of magnitude smaller that the oscillation persiod of the
plasmon wave can induce time reversal. We find that for all pulse
durations $\Delta t$, time reversal is clearly observable (see Fig.
\ref{fig:time_reversal_vs_pulse_duration}), although the wavefront
is distorted for $\Delta t=0.5T_{\mathrm{pl}}$. We conclude that
pulse induced time reversal is quite robust to the duration of the
pulse as compared to the duration of one oscillation cycle of the
plasmon wave, and that time inversion is observable for a pulse which
is half as long as the oscillation cycle. In fact, in recent experiments
the time reversal of optical signals was measured for pulse durations
comparable to the duration of one optical cycle .

\subsection{Estimating the critical driving strength \label{subsec:Estimating-the-critical_driving}}

We assume a gap size of $E_{g}=0.3\,\mathrm{eV}$, a Floquet driving
frequency of $\hbar\Omega_{F}=0.35\,\mathrm{eV}$ and a pseudospin-orbit
coupling of $\lambda=15\,\mathrm{eV}\text{Å}$. An electron density
of $\bar{\rho}=1.18\cdot10^{11}/\mathrm{cm}^{-2}$ guarantees that
the Fermi energy is close to, but above the resonance. The plasmon
quality factor $Q$ sets a lower bound on the required driving and
modulation strength. Assuming $Q=\omega_{1}/\gamma\approx10^{2}$,
the instability condition $h>2\gamma/\omega_{1}$ yields a critical
driving strength of $\mathcal{\bar{E}}\approx4\cdot10^{5}\,\mathrm{V}/\mathrm{m}$
with a modulation amplitude $\delta\mathcal{\bar{E}}=0.5\mathcal{\bar{E}}$.
Such field strengths can be reached with continuous wave lasers. The
corresponding intensity is by a factor of $10^{4}$ smaller than in
current solid state Floquet engineering experiments \citep{wang2013_floquet-bloch_states_observation,mahmood2016selective_scattering_floquet-bloch_volkov,mciver2020light_anomaouls_hall_graphene,zhou2023black_phosphorus_floquet}.

\subsection{Heating and Landau damping \label{subsec:Heating-and-Landau}}

In undoped gapped systems, the main heating mechanisms are radiative
recombination and momentum conserving single photon absorption \citep{seetharam2015baths_controlled_floquet_population,esin2021_liquid_crystal}.
In the proposed set-up, these processes are suppressed by the exclusion
principle. The most relevant allowed heating processes are phonon-
or disorder-assisted, momentum non-conserving single-photon absorption
and interaction-assisted single-photon absorption. Here we estimate
the heating due to these processes and show that the Landau damping
induced by this heating is very small.

In the following we calculate the number of electrons excited by momentum
non-conserving single-photon absorptions per unit of time and area
$\Gamma_{\gamma}$, and the corresponding number for interaction-aided
single-photon absorptions $\Gamma_{\mathrm{int}}$.

\subsubsection{Estimation of $\Gamma_{\gamma}$}

The matrix element for the absorption of n photons is proportional
to $\left(e\mathcal{E}\lambda/\hbar\Omega_{F}^{2}\right)^{n}$ (see,
e.g., \citep{esin2021_liquid_crystal}) and the rate of momentum non-conserving
scattering is $\gamma$. Thus, according to Fermi's Golden rule, the
scattering rate can be estimated as $\gamma\left(e\mathcal{E}\lambda/\hbar\Omega_{F}^{2}\right)^{2n}$.
To find $\Gamma_{\gamma}$, we have to multiply by the density of
electrons available for this kind of scattering. Slightly overestimating
this density (and the heating due to this process), we approximate
it by the total electron density of the upper band $\rho=\pi k_{F}^{2}/\left(2\pi\right)^{2}$.
We thus find
\begin{equation}
\Gamma_{\gamma}\approx\frac{\gamma\pi k_{F}^{2}}{\left(2\pi\right)^{2}}\left(\frac{e\mathcal{E}\lambda}{\hbar\Omega_{F}^{2}}\right)^{2}.\label{eq:rate1}
\end{equation}
Using the estimates from the main text ($\hbar\Omega_{F}=0.35\,\mathrm{eV}$,
$\lambda=15\,\mathrm{eV}\text{Å}$, $\mathcal{\bar{E}}\approx4\cdot10^{5}\,\mathrm{V}/\mathrm{m}$),
the dimensionless amplitude of the electric field controlling the
strength of the photon absorption is
\[
\frac{e\mathcal{E}\lambda}{\hbar\Omega_{F}^{2}}\approx5\cdot10^{-3}.
\]
We find
\begin{align*}
\Gamma_{\gamma} & \approx2.5\cdot10^{-5}\cdot\gamma\rho\\
 & \approx1.6\cdot10^{6}\mathrm{Hz}\cdot\rho\cdot
\end{align*}
We assumed $\gamma\approx6.28\cdot10^{10}\mathrm{Hz}$ for a quality
factor $Q=\omega/\gamma=10^{2}$ and a plasmon frequency of $\omega/2\pi=1\,\mathrm{THz}$.

\subsubsection{Estimation of $\Gamma_{\mathrm{int}}$}

In this process, a single photon is absorbed and its energy is distributed
between two electrons. The two electrons exchange a momentum $\mathbf{q}$
through Coulomb interaction. Energy conservation yields
\[
\varepsilon\left(\mathbf{k}_{1}+\mathbf{q}\right)+\varepsilon\left(\mathbf{k}_{2}-\mathbf{q}\right)-\varepsilon\left(\mathbf{k}_{2}\right)-\varepsilon\left(\mathbf{k}_{1}\right)-\hbar\Omega_{F}=0.
\]
Thus, the available phase space for this process is very limited.
The Coulomb potential is given by
\[
V\left(\mathbf{q}\right)=\frac{e^{2}}{\varepsilon}\frac{2\pi}{q+q_{s}},
\]
where the screening length is $q_{s}\approx4\alpha k_{F}$, with the
fine structure constant $\alpha=e^{2}/\varepsilon\lambda$ \citep{pertsova2018excitonic_instability}.
This approximation is valid when the Fermi energy is far enough from
the bottom of the band and the dispersion is nearly linear. For the
estimate at hand, we approximate
\begin{align*}
V\left(0\right) & \approx\frac{e^{2}}{\varepsilon}\frac{2\pi}{q_{s}}\\
\varepsilon\left(\mathbf{k}\right) & \approx\lambda k
\end{align*}
The number of absorbed photons per unit of time can then be estimated
using Fermi's Golden rule as
\[
\Gamma_{\mathrm{int}}\approx\frac{1}{\hbar}\int\frac{d^{2}k_{1}}{\left(2\pi\right)^{2}}\int\frac{d^{2}k_{2}}{\left(2\pi\right)^{2}}\int\frac{d^{2}q}{\left(2\pi\right)^{2}}\delta\left(\varepsilon\left(\mathbf{k}_{1}+\mathbf{q}\right)+\varepsilon\left(\mathbf{k}_{2}-\mathbf{q}\right)-\varepsilon\left(\mathbf{k}_{2}\right)-\varepsilon\left(\mathbf{k}_{1}\right)-\hbar\Omega_{F}\right)V^{2}\left(0\right)\left(\frac{e\mathcal{E}\lambda}{\hbar\Omega_{F}^{2}}\right)^{2}.
\]
To take the integrals, it is useful to introduce dimensionless variables
\begin{align*}
\mathbf{K}_{i} & =\mathbf{k}_{i}/k_{F}\\
\mathbf{Q} & =\mathbf{q}/k_{F}\\
\bar{\Omega} & =\hbar\Omega_{F}/\lambda k_{F}
\end{align*}
The above integral can then be written as 
\[
\Gamma_{\mathrm{int}}=\frac{V^{2}\left(0\right)k_{F}^{6}}{\lambda\hbar k_{F}}\left(\frac{e\mathcal{E}\lambda}{\hbar\Omega_{F}^{2}}\right)^{2}\int\frac{d^{2}K_{1}}{\left(2\pi\right)^{2}}\int\frac{d^{2}K_{2}}{\left(2\pi\right)^{2}}\int\frac{d^{2}Q}{\left(2\pi\right)^{2}}\delta\left(\left|\mathbf{K}_{1}+\mathbf{Q}\right|+\left|\mathbf{K}_{2}-\mathbf{Q}\right|-\left|\mathbf{K}_{2}\right|-\left|\mathbf{K}_{1}\right|-\bar{\Omega}\right).
\]
The $q$ integration can be performed by going to elliptical coordinates.
This procedure (for details see e.g. \citep{sachdev1998_elliptic_coordinates})
yields
\[
\Gamma_{\mathrm{int}}=\frac{V^{2}\left(0\right)k_{F}^{5}}{\hbar\lambda}\left(\frac{e\mathcal{E}\lambda}{\hbar\Omega_{F}^{2}}\right)^{2}\int\frac{d^{2}K_{1}}{\left(2\pi\right)^{2}}\int\frac{d^{2}K_{2}}{\left(2\pi\right)^{2}}\int\frac{d\vartheta}{\left(2\pi\right)^{2}}\left|\mathbf{K}_{1}+\mathbf{K}_{2}\right|\frac{\left(\frac{K_{1}+K_{2}+\bar{\Omega}}{\left|\mathbf{K}_{1}+\mathbf{K}_{2}\right|}\right)^{2}-\cos^{2}\vartheta}{4\sqrt{\left(\frac{K_{1}+K_{2}+\bar{\Omega}}{\left|\mathbf{K}_{1}+\mathbf{K}_{2}\right|}\right)^{2}-1}}.
\]
For the parameters of our estimate ($\hbar\Omega=0.35\,\mathrm{eV}$,
$\lambda=15\,\mathrm{eV}\text{Å}$, $\mathcal{\bar{E}}\approx4\cdot10^{5}\,\mathrm{V}/\mathrm{m}$,
$k_{F}=1.2\cdot10^{8}\,\mathrm{m^{-1}}$), we find
\[
\bar{\Omega}=1.9.
\]
A numerical evaluation of the remaining integral gives
\begin{align*}
\Gamma_{\mathrm{int}} & \approx1.02\cdot10^{-3}\cdot\frac{V^{2}\left(0\right)k_{F}^{5}}{\hbar\lambda}\left(\frac{e\mathcal{E}\lambda}{\hbar\Omega_{F}^{2}}\right)^{2}\\
 & =1.02\cdot10^{-3}\cdot\frac{V^{2}\left(0\right)k_{F}^{5}}{\hbar\lambda}\left(\frac{e\mathcal{E}\lambda}{\hbar\Omega_{F}^{2}}\right)^{2}
\end{align*}
On the other hand
\[
V\left(0\right)\approx\frac{e^{2}\pi}{2\varepsilon\alpha k_{F}}.
\]
Collecting everything and assuming $\varepsilon=6\varepsilon_{0}$,
we find (SI Units)
\begin{align*}
\Gamma_{\mathrm{int}} & \approx1.02\cdot10^{-3}\cdot\frac{\pi e^{4}k_{F}^{3}}{16\hbar\varepsilon^{2}\alpha^{2}\lambda}\left(\frac{e\mathcal{E}\lambda}{\hbar\Omega_{F}^{2}}\right)^{2}\\
 & \approx6.1\cdot10^{6}\,\mathrm{Hz}\cdot\rho,
\end{align*}
where we used our estimate
\[
\frac{e\mathcal{E}\lambda}{\hbar\Omega_{F}^{2}}\approx5\cdot10^{-3}.
\]

\subsubsection{Landau damping}

The Landau damping of plasmons is described by the imaginary part
of the Lindhard function
\[
\chi\left(\omega,\mathbf{q}\right)=\int\frac{d^{2}k}{\left(2\pi\right)^{2}}\frac{f\left(\varepsilon_{\mathbf{k}-\mathbf{q}/2}\right)-f\left(\varepsilon_{\mathbf{k}+\mathbf{q}/2}\right)}{\omega+i0^{+}+\varepsilon_{\mathbf{k}-\mathbf{q}/2}-\varepsilon_{\mathbf{k}+\mathbf{q}/2}}.
\]
The imaginary part is
\[
\mathrm{Im}\left[\chi\left(\omega,\mathbf{q}\right)\right]=-\pi\int\frac{kdk}{\left(2\pi\right)^{2}}\int d\varphi\left[f\left(\varepsilon_{\mathbf{k}}\right)-f\left(\varepsilon_{\mathbf{k}+\mathbf{q}}\right)\right]\delta\left(\omega+\varepsilon_{\mathbf{k}}-\varepsilon_{\mathbf{k}+\mathbf{q}}\right).
\]
Let $\varepsilon_{\mathbf{k}}=k^{2}/2m^{*}$. This assumption simplifies
the calculations but can be made without loss of generality, because
in the end we only need to consider the linearized dispersion in the
vicinity of the Fermi surface. Defining the quantities $\mathbf{K}=\mathbf{k}/\sqrt{2m^{*}}$,
$\mathbf{Q}'=\mathbf{q}/\sqrt{2m^{*}}$, we have
\begin{align*}
\mathrm{Im}\left[\chi\left(\omega,\mathbf{q}\right)\right] & =-\pi\int\frac{kdk}{\left(2\pi\right)^{2}}\int d\varphi\left[f\left(K^{2}\right)-f\left(K^{2}+Q'^{2}+2KQ'\cos\varphi\right)\right]\delta\left(Q'^{2}+2KQ'\cos\varphi-\omega\right)\\
 & \quad-2m^{*}\pi\int\frac{KdK}{\left(2\pi\right)^{2}}\int d\varphi\left[f\left(K^{2}\right)-f\left(K^{2}+Q'^{2}+2KQ'\cos\varphi\right)\right]\delta\left(\frac{Q'^{2}-\omega}{2KQ'}+\cos\varphi\right)
\end{align*}
We are interested in the case $\frac{Q'^{2}-\omega}{2KQ'}<0$, i.e.
$-\pi/2<\varphi<\pi/2$. We substitute
\begin{align*}
\mu & =\cos\varphi\\
d\varphi & =-\frac{d\mu}{\sin\varphi}=-\frac{d\mu}{\pm\sqrt{1-\cos^{2}\varphi}}
\end{align*}
where the plus-minus-sign indicates that we have to distinguish the
cases $\varphi\lessgtr0$. We find
\begin{align}
\mathrm{Im}\left[\chi\left(\omega,\mathbf{Q}'\right)\right] & =-\frac{m^{*}\pi}{Q'}\int\frac{dK}{\left(2\pi\right)^{2}}\int_{-1}^{1}\frac{d\mu}{\sqrt{1-\mu^{2}}}\left[f\left(K^{2}\right)-f\left(K^{2}+Q'^{2}+2KQ'\mu\right)\right]\delta\left(\frac{Q'^{2}-\omega}{2KQ'}+\mu\right)\nonumber \\
 & =-\frac{m^{*}\pi}{Q'}\int_{\frac{\omega-Q^{2}}{2Q}}^{\infty}\frac{dK}{\left(2\pi\right)^{2}}\frac{1}{\sqrt{1-\left(\frac{Q'^{2}-\omega}{2KQ'}\right)^{2}}}\left[f\left(K^{2}\right)-f\left(K^{2}+\omega\right)\right],\label{eq:int_to_appr}
\end{align}
where the last step follows from 
\[
-1<\frac{Q'^{2}-\omega}{2KQ'}<0
\]
For $T=0$, approximating $f\left(K^{2}\right)-f\left(K^{2}-\omega\right)\approx-\delta\left(K-K_{F}\right)\omega/K_{F}$,
we find
\begin{align*}
\mathrm{Im}\left[\chi\left(\omega,\mathbf{q}\right)\right] & =\frac{-m^{*}\omega}{4\pi QK_{F}}\frac{\Theta\left(K_{F}-\frac{Q'^{2}+\omega}{2Q}\right)}{\sqrt{1-\frac{\left(Q'^{2}-\omega\right)^{2}}{4K_{F}^{2}Q'^{2}}}}
\end{align*}
As the theta function indicates, plasmons are undamped for
\[
\frac{qk_{F}}{m^{*}}<\omega-\frac{q^{2}}{2m}.
\]
In our case $\omega\gg q^{2}/2m^{*}$ and the condition reduces to
\[
\omega>v_{F}q.
\]

At finite temperatures and with $\omega/Q'\gg Q'$, the integral in
Eq. (\ref{eq:int_to_appr}) can be approximated as
\begin{align*}
\mathrm{Im}\left[\chi\left(\omega,\mathbf{q}\right)\right] & \approx-\frac{m^{*}\pi}{Q'}\int_{\frac{\omega}{2Q}}^{\infty}\frac{dK}{\left(2\pi\right)^{2}}\frac{K}{\sqrt{K^{2}-\left(\frac{\omega}{2Q'}\right)^{2}}}\left[f\left(K^{2}\right)-f\left(K^{2}+\omega\right)\right]
\end{align*}
For $\omega\gg v_{F}q=K_{F}Q'/2$, we can approximate the Fermi-Dirac
distributions as exponentials
\begin{align}
\mathrm{Im}\left[\chi\left(\omega,\mathbf{q}\right)\right] & \approx-\frac{m^{*}\pi}{Q'}\int_{\frac{\omega}{2Q}}^{\infty}\frac{dK}{\left(2\pi\right)^{2}}\frac{K}{\sqrt{K^{2}-\left(\frac{\omega}{2Q}\right)^{2}}}\left[e^{-\beta K^{2}}-e^{-\beta\left(K^{2}+\omega\right)}\right]\nonumber \\
 & =-\frac{m^{*}\pi}{\left(2\pi\right)^{2}Q'}\int_{\left(\frac{\omega}{2Q}\right)^{2}}^{\infty}dE\frac{1}{\sqrt{E-\left(\frac{\omega}{2Q'}\right)^{2}}}\left[e^{-\beta E}-e^{-\beta\left(E+\omega\right)}\right]\nonumber \\
 & \approx-\frac{m^{*}\omega}{4\pi Q'}\int_{\left(\frac{\omega}{2Q}\right)^{2}}^{\infty}dE\frac{\beta e^{-\beta E}}{\sqrt{E-\left(\frac{\omega}{2Q'}\right)^{2}}}=-\frac{m^{*}\omega}{4\pi Q'}\sqrt{\beta}\int_{\beta\left(\frac{\omega}{2Q}\right)^{2}}^{\infty}dx\frac{e^{-x}}{\sqrt{x-\beta\left(\frac{\omega}{2Q'}\right)^{2}}}\nonumber \\
 & =-\frac{m^{*}\omega}{4\pi Q'}\sqrt{\beta}e^{-\beta\left(\frac{\omega}{2Q'}\right)^{2}}\int_{0}^{\infty}dx\frac{e^{-x}}{\sqrt{x}}\nonumber \\
 & =-\frac{m^{*}\omega\sqrt{\beta}}{4\sqrt{\pi}Q'}e^{-\beta\left(\frac{\omega}{2Q'}\right)^{2}}\label{eq:Imag_lindhard}
\end{align}

\subsubsection{Temperature estimate}

We estimate the effective temperature of electrons using experimental
data on the cooling power provided by phonons. In a Floquet driven
system, the effective temperature of the electrons and the temperature
of the lattice, which is coupled to a cold bath, can differ. The cooling
power is the amount of energy per unit of time that is carried away
by the cold lattice (mostly by optical phonons), thus stabilizing
the effective electron temperature. While experimental results for
the cooling power are available for graphene \citep{baker2013cooling_power_graphene_exp2,baker2012_cooling_power_graphene_exp1,betz2012_cooling_power_graphene_exp_values},
theoretical work \citep{kaasbjerg2014_cooling_power_tmds} shows that
the cooling power of TMDs is even larger, leading to a lower effective
temperature. Here, for a conservative estimate, we use the values
measured for graphene \citep{betz2012_cooling_power_graphene_exp_values},
where the lattice was kept at $4.2\,\mathrm{K}$.

The cooling power per electron $P$ strongly depends on the effective
temperature of the electrons. In two dimensional materials \citep{kubakaddi2009_cooling_power_graphene_theory,kaasbjerg2014_cooling_power_tmds}
one finds
\[
P_{\mathrm{cool}}=\frac{\Sigma\left(\bar{\rho}\right)}{\bar{\rho}}\left(T_{e}^{4}-T_{\mathrm{ph}}^{4}\right).
\]
This law holds well for electron temperatures below $100\,\mathrm{K}$.
$T_{e}$ is the effective electron temperature, and $T_{\mathrm{ph}}$
is the temperature of the lattice. The above assumes that the electrons
are thermalized. In fact, one expects a very fast thermalization of
the electrons due to electron-electron interactions. For the above
estimate of $\bar{\rho}=1.18\cdot10^{11}/\mathrm{cm}^{-2}$, we find
$\Sigma\left(\bar{\rho}\right)\approx1\frac{\mathrm{mW}}{\mathrm{K}^{4}\mathrm{m}^{2}}$.
On the other hand, the power supplied by Floquet drive is 
\begin{align*}
P_{\mathrm{drive}} & \approx\hbar\Omega_{F}\left(\Gamma_{\mathrm{int}}+\Gamma_{\gamma}\right)/\bar{\rho}.
\end{align*}
Setting $P_{\mathrm{cool}}=P_{\mathrm{drive}}$, we find that cooling
and heating rates are balanced at the effective electron temperature
\begin{equation}
T_{e}\approx20\,\mathrm{K}.\label{eq:Temperature_estimate}
\end{equation}

Eq. (\ref{eq:Temperature_estimate}) shows that the heating induced
by MFPD is not too large. In fact, due to the nonlinear dependence
of the cooling power on $T_{e}$, the driving power could be easily
increased by three orders of magnitude giving a feasible effective
temperature of around $T_{e}\approx100\,\mathrm{K}.$

Let us finally comment on the Landau damping induced by MFPD. To estimate
the factor $e^{-\beta\left(\frac{\omega}{2Q}\right)^{2}}$ in Eq.
(\ref{eq:Imag_lindhard}), we assume $T=300\,K$ and neglect the small
MFPD induced heating. Typically, in experiments, the temperature as
well as the Landau damping will be smaller. For a typical plasmon
with wavelength $1\upmu m$ and roughly $\omega=1\,\mathrm{THz}$,
we find 
\[
\beta\left(\frac{\sqrt{m^{*}}\omega}{2q}\right)^{2}\sim10^{3},
\]
such that even at room temperature, the Landau damping is suppressed
by a factor of 
\[
\sim e^{-10^{3}}.
\]


\begin{thebibliography}{10}

\bibitem{galiffi2022photonics_time_varying_ptc}
E.~Galiffi, R.~Tirole, S.~Yin, H.~Li, S.~Vezzoli, P.~A. Huidobro, M.~G.
  Silveirinha, R.~Sapienza, A.~Al{\`u}, and J.~Pendry, \textit{Photonics of
  time-varying media}, Advanced Photonics \textbf{4}, 014002 (2022).

\bibitem{morgenthaler1958velocity_modulated}
F.~R. Morgenthaler, \textit{Velocity modulation of electromagnetic waves}, IRE
  Transactions on microwave theory and techniques \textbf{6}, 167 (1958).

\bibitem{holberg1966parametric_PTC}
D.~Holberg and K.~Kunz, \textit{Parametric properties of fields in a slab of
  time-varying permittivity}, IEEE Transactions on Antennas and Propagation
  \textbf{14}, 183 (1966).

\bibitem{tirole2022_time_varying_mirror}
R.~Tirole, E.~Galiffi, J.~Dranczewski, T.~Attavar, B.~Tilmann, Y.-T. Wang,
  P.~A. Huidobro, A.~Al{\'u}, J.~B. Pendry, S.~A. Maier \textit{et~al.},
  \textit{Saturable time-varying mirror based on an epsilon-near-zero
  material}, Physical Review Applied \textbf{18}, 054067 (2022).

\bibitem{peng2022topological_spacetime}
Y.~Peng, \textit{Topological space-time crystal}, Physical Review Letters
  \textbf{128}, 186802 (2022).

\bibitem{sharabi2022spatiotemporal_PTC}
Y.~Sharabi, A.~Dikopoltsev, E.~Lustig, Y.~Lumer, and M.~Segev,
  \textit{Spatiotemporal photonic crystals}, Optica \textbf{9}, 585 (2022).

\bibitem{wang2023metasurface_PTC}
X.~Wang, M.~S. Mirmoosa, V.~S. Asadchy, C.~Rockstuhl, S.~Fan, and S.~A.
  Tretyakov, \textit{Metasurface-based realization of photonic time crystals},
  Science Advances \textbf{9}, eadg7541 (2023).

\bibitem{wang2023controlling_surface_waves}
X.~Wang, M.~S. Mirmoosa, and S.~A. Tretyakov, \textit{Controlling surface waves
  with temporal discontinuities of metasurfaces}, Nanophotonics  (2023).

\bibitem{koutserimpas2023multiharmonic_time_varied_metasurfaces}
T.~T. Koutserimpas and C.~Valagiannopoulos, \textit{Multiharmonic resonances of
  coupled time-modulated resistive metasurfaces}, Physical Review Applied
  \textbf{19}, 064072 (2023).

\bibitem{akbarzadeh2018_temporal_switch}
A.~Akbarzadeh, N.~Chamanara, and C.~Caloz, \textit{Inverse prism based on
  temporal discontinuity and spatial dispersion}, Optics letters \textbf{43},
  3297 (2018).

\bibitem{pacheco2020temporal_switching}
V.~Pacheco-Pe{\~n}a and N.~Engheta, \textit{Temporal aiming}, Light: Science \&
  Applications \textbf{9}, 129 (2020).

\bibitem{pacheco2021temporal_switching2}
V.~Pacheco-Pe{\~n}a and N.~Engheta, \textit{Temporal equivalent of the brewster
  angle}, Physical Review B \textbf{104}, 214308 (2021).

\bibitem{lyubarov2022photonic_time_crystal_amplified}
M.~Lyubarov, Y.~Lumer, A.~Dikopoltsev, E.~Lustig, Y.~Sharabi, and M.~Segev,
  \textit{Amplified emission and lasing in photonic time crystals}, Science
  \textbf{377}, 425 (2022).

\bibitem{mendoncca2005_entangled_photon_pairs}
J.~Mendon{\c{c}}a and A.~Guerreiro, \textit{Time refraction and the quantum
  properties of vacuum}, Physical Review A \textbf{72}, 063805 (2005).

\bibitem{lustig2018topological_photonic_time_crystal}
E.~Lustig, Y.~Sharabi, and M.~Segev, \textit{Topological aspects of photonic
  time crystals}, Optica \textbf{5}, 1390 (2018).

\bibitem{carminati2021universal_statistics_time-varying_medium}
R.~Carminati, H.~Chen, R.~Pierrat, and B.~Shapiro, \textit{Universal statistics
  of waves in a random time-varying medium}, Physical Review Letters
  \textbf{127}, 094101 (2021).

\bibitem{wang2018photonic_non_hermitian}
N.~Wang, Z.-Q. Zhang, and C.~T. Chan, \textit{Photonic floquet media with a
  complex time-periodic permittivity}, Physical Review B \textbf{98}, 085142
  (2018).

\bibitem{li2021_non_hermition_TPT}
H.~Li, S.~Yin, E.~Galiffi, and A.~Al{\`u}, \textit{Temporal parity-time
  symmetry for extreme energy transformations}, Physical Review Letters
  \textbf{127}, 153903 (2021).

\bibitem{pan2022superluminal_k-gap_solitons}
Y.~Pan, M.-I. Cohen, and M.~Segev, \textit{Superluminal k-gap solitons in
  photonic time-crystals with kerr nonlinearity}, in \textit{CLEO:
  QELS\_Fundamental Science}, Optica Publishing Group (2022), (FW5J--5).

\bibitem{bacot2016time_water}
V.~Bacot, M.~Labousse, A.~Eddi, M.~Fink, and E.~Fort, \textit{Time reversal and
  holography with spacetime transformations}, Nature Physics \textbf{12}, 972
  (2016).

\bibitem{fleury2016_topological_sound}
R.~Fleury, A.~B. Khanikaev, and A.~Alu, \textit{Floquet topological insulators
  for sound}, Nature communications \textbf{7}, 11744 (2016).

\bibitem{wen2022_acoustic_non_hermitian}
X.~Wen, X.~Zhu, A.~Fan, W.~Y. Tam, J.~Zhu, H.~W. Wu, F.~Lemoult, M.~Fink, and
  J.~Li, \textit{Unidirectional amplification with acoustic non-hermitian
  space- time varying metamaterial}, Communications Physics \textbf{5}, 18
  (2022).

\bibitem{oka2009photovoltaic_hall_effect}
T.~Oka and H.~Aoki, \textit{Photovoltaic hall effect in graphene}, Physical
  Review B \textbf{79}, 081406 (2009).

\bibitem{kitagawa2011floquetinduced}
T.~Kitagawa, T.~Oka, A.~Brataas, L.~Fu, and E.~Demler, \textit{Transport
  properties of nonequilibrium systems under the application of light:
  Photoinduced quantum hall insulators without landau levels}, Physical Review
  B \textbf{84}, 235108 (2011).

\bibitem{kitagawa2010topological_characterization_driven_quantum_system}
T.~Kitagawa, E.~Berg, M.~Rudner, and E.~Demler, \textit{Topological
  characterization of periodically driven quantum systems}, Physical Review B
  \textbf{82}, 235114 (2010).

\bibitem{lindner2011floquet}
N.~H. Lindner, G.~Refael, and V.~Galitski, \textit{Floquet topological
  insulator in semiconductor quantum wells}, Nature Physics \textbf{7}, 490
  (2011).

\bibitem{wang2013_floquet-bloch_states_observation}
Y.~Wang, H.~Steinberg, P.~Jarillo-Herrero, and N.~Gedik, \textit{Observation of
  {F}loquet-{B}loch states on the surface of a topological insulator}, Science
  \textbf{342}, 453 (2013).

\bibitem{mciver2020light_anomaouls_hall_graphene}
J.~W. McIver, B.~Schulte, F.-U. Stein, T.~Matsuyama, G.~Jotzu, G.~Meier, and
  A.~Cavalleri, \textit{Light-induced anomalous hall effect in graphene},
  Nature physics \textbf{16}, 38 (2020).

\bibitem{mahmood2016selective_scattering_floquet-bloch_volkov}
F.~Mahmood, C.-K. Chan, Z.~Alpichshev, D.~Gardner, Y.~Lee, P.~A. Lee, and
  N.~Gedik, \textit{Selective scattering between {F}loquet--{B}loch and
  {V}olkov states in a topological insulator}, Nature Physics \textbf{12}, 306
  (2016).

\bibitem{zhou2023black_phosphorus_floquet}
S.~Zhou, C.~Bao, B.~Fan, H.~Zhou, Q.~Gao, H.~Zhong, T.~Lin, H.~Liu, P.~Yu,
  P.~Tang \textit{et~al.}, \textit{Pseudospin-selective {F}loquet band
  engineering in black phosphorus}, Nature \textbf{614}, 75 (2023).

\bibitem{usaj2014_floquet_graphene_topo}
G.~Usaj, P.~M. Perez-Piskunow, L.~F. Torres, and C.~A. Balseiro,
  \textit{Irradiated graphene as a tunable floquet topological insulator},
  Physical Review B \textbf{90}, 115423 (2014).

\bibitem{perez2014floquet_traphene_topo}
P.~M. Perez-Piskunow, G.~Usaj, C.~A. Balseiro, and L.~F. Torres,
  \textit{Floquet chiral edge states in graphene}, Physical Review B
  \textbf{89}, 121401 (2014).

\bibitem{oka2019floquet_review}
T.~Oka and S.~Kitamura, \textit{Floquet engineering of quantum materials},
  Annual Review of Condensed Matter Physics \textbf{10}, 387 (2019).

\bibitem{katz2020optically}
O.~Katz, G.~Refael, and N.~H. Lindner, \textit{Optically induced flat bands in
  twisted bilayer graphene}, Physical Review B \textbf{102}, 155123 (2020).

\bibitem{castro2022optimal_floquet_control}
A.~Castro, U.~De~Giovannini, S.~A. Sato, H.~H{\"u}bener, and A.~Rubio,
  \textit{Floquet engineering the band structure of materials with optimal
  control theory}, Physical Review Research \textbf{4}, 033213 (2022).

\bibitem{esin2018q_steady_state_topo_ins}
I.~Esin, M.~S. Rudner, G.~Refael, and N.~H. Lindner, \textit{Quantized
  transport and steady states of floquet topological insulators}, Physical
  Review B \textbf{97}, 245401 (2018).

\bibitem{esin2020floquet_metal_insulator}
I.~Esin, M.~S. Rudner, and N.~H. Lindner, \textit{Floquet metal-to-insulator
  phase transitions in semiconductor nanowires}, Science advances \textbf{6},
  eaay4922 (2020).

\bibitem{esin2021_liquid_crystal}
I.~Esin, G.~K. Gupta, E.~Berg, M.~S. Rudner, and N.~H. Lindner,
  \textit{Electronic floquet gyro-liquid crystal}, Nature communications
  \textbf{12}, 1 (2021).

\bibitem{dehghani2015_floquet_topo}
H.~Dehghani, T.~Oka, and A.~Mitra, \textit{Out-of-equilibrium electrons and the
  hall conductance of a floquet topological insulator}, Physical Review B
  \textbf{91}, 155422 (2015).

\bibitem{genske2015floquet_boltzmann}
M.~Genske and A.~Rosch, \textit{Floquet-boltzmann equation for periodically
  driven fermi systems}, Physical Review A \textbf{92}, 062108 (2015).

\bibitem{glazman1983kinetics_pulses_semiconductor}
L.~Glazman, \textit{Kinetics of electrons and holes in direct-gap
  semiconductors photo-excited by high-intensity pulses}, Soviet Physics
  Semiconductors-USSR \textbf{17}, 494 (1983).

\bibitem{dehghani2014dissipative_topo_floquet}
H.~Dehghani, T.~Oka, and A.~Mitra, \textit{Dissipative floquet topological
  systems}, Physical Review B \textbf{90}, 195429 (2014).

\bibitem{sentef2015pump_probe_floquet}
M.~Sentef, M.~Claassen, A.~Kemper, B.~Moritz, T.~Oka, J.~Freericks, and
  T.~Devereaux, \textit{Theory of floquet band formation and local pseudospin
  textures in pump-probe photoemission of graphene}, Nature communications
  \textbf{6}, 7047 (2015).

\bibitem{chan2016floquet_ref}
C.-K. Chan, P.~A. Lee, K.~S. Burch, J.~H. Han, and Y.~Ran, \textit{When chiral
  photons meet chiral fermions: photoinduced anomalous hall effects in weyl
  semimetals}, Physical review letters \textbf{116}, 026805 (2016).

\bibitem{farrell2015floquet_ref}
A.~Farrell and T.~Pereg-Barnea, \textit{Photon-inhibited topological transport
  in quantum well heterostructures}, Physical Review Letters \textbf{115},
  106403 (2015).

\bibitem{gu2011floquet_ref}
Z.~Gu, H.~Fertig, D.~P. Arovas, and A.~Auerbach, \textit{Floquet spectrum and
  transport through an irradiated graphene ribbon}, Physical review letters
  \textbf{107}, 216601 (2011).

\bibitem{hubener2017floquet_ref}
H.~H{\"u}bener, M.~A. Sentef, U.~De~Giovannini, A.~F. Kemper, and A.~Rubio,
  \textit{Creating stable floquet--weyl semimetals by laser-driving of 3d
  {D}irac materials}, Nature communications \textbf{8}, 13940 (2017).

\bibitem{jiang2011floquet_ref}
L.~Jiang, T.~Kitagawa, J.~Alicea, A.~Akhmerov, D.~Pekker, G.~Refael, J.~I.
  Cirac, E.~Demler, M.~D. Lukin, and P.~Zoller, \textit{Majorana fermions in
  equilibrium and in driven cold-atom quantum wires}, Physical review letters
  \textbf{106}, 220402 (2011).

\bibitem{kennes2019floquet_ref}
D.~M. Kennes, N.~M{\"u}ller, M.~Pletyukhov, C.~Weber, C.~Bruder, F.~Hassler,
  J.~Klinovaja, D.~Loss, and H.~Schoeller, \textit{Chiral one-dimensional
  floquet topological insulators beyond the rotating wave approximation},
  Physical Review B \textbf{100}, 041103 (2019).

\bibitem{kundu2013floquet_ref}
A.~Kundu and B.~Seradjeh, \textit{Transport signatures of floquet majorana
  fermions in driven topological superconductors}, Physical review letters
  \textbf{111}, 136402 (2013).

\bibitem{thakurathi2017floquet_ref}
M.~Thakurathi, D.~Loss, and J.~Klinovaja, \textit{Floquet majorana fermions and
  parafermions in driven rashba nanowires}, Physical Review B \textbf{95},
  155407 (2017).

\bibitem{kiselev2023MFPD}
E.~I. Kiselev, M.~S. Rudner, and N.~H. Lindner, \textit{Modulated floquet
  parametric driving and non-equilibrium crystalline electron states}, arXiv
  preprint arXiv:2303.02148  (2023).

\bibitem{rezende2019introduction_antiferromagnetic_magnons}
S.~M. Rezende, A.~Azevedo, and R.~L. Rodr{\'\i}guez-Su{\'a}rez,
  \textit{Introduction to antiferromagnetic magnons}, Journal of Applied
  Physics \textbf{126} (2019).

\bibitem{kreisel2009_YIG_magnon_spectra}
A.~Kreisel, F.~Sauli, L.~Bartosch, and P.~Kopietz, \textit{Microscopic
  spin-wave theory for yttrium-iron garnet films}, The European physical
  journal B \textbf{71}, 59 (2009).

\bibitem{mentink2015_Mott_exchange_floquet}
J.~Mentink, K.~Balzer, and M.~Eckstein, \textit{Ultrafast and reversible
  control of the exchange interaction in {M}ott insulators}, Nature
  communications \textbf{6}, 6708 (2015).

\bibitem{chaudhary2019orbital_floquet_engineering_exchange}
S.~Chaudhary, D.~Hsieh, and G.~Refael, \textit{Orbital floquet engineering of
  exchange interactions in magnetic materials}, Physical Review B \textbf{100},
  220403 (2019).

\bibitem{ron2020_light_induced_enhancement_of_exchange}
A.~Ron, S.~Chaudhary, G.~Zhang, H.~Ning, E.~Zoghlin, S.~Wilson, R.~Averitt,
  G.~Refael, and D.~Hsieh, \textit{Ultrafast enhancement of ferromagnetic spin
  exchange induced by ligand-to-metal charge transfer}, Physical Review Letters
  \textbf{125}, 197203 (2020).

\bibitem{lundeberg2017_koppens_plasmonics_near_field}
M.~B. Lundeberg, Y.~Gao, R.~Asgari, C.~Tan, B.~Van~Duppen, M.~Autore,
  P.~Alonso-Gonz{\'a}lez, A.~Woessner, K.~Watanabe, T.~Taniguchi,
  R.~Hillenbrand, J.~Hone, M.~Polini, and F.~H.~L. Koppens, \textit{Tuning
  quantum nonlocal effects in graphene plasmonics}, Science \textbf{357}, 187
  (2017).

\bibitem{landau_lifshitz_mechanics}
L.~D. Landau and E.~M. Lifshitz, \textit{Mechanics}, Butterworth-Heinemann
  (1976). ISBN 978-0750628969.

\bibitem{turyn1993_mathieu_threshold_high_m}
L.~Turyn, \textit{The damped {M}athieu equation}, Quarterly of applied
  mathematics \textbf{51}, 389 (1993).

\bibitem{zurita2009reflection_epsilon_time_dep}
J.~R. Zurita-S{\'a}nchez, P.~Halevi, and J.~C. Cervantes-Gonzalez,
  \textit{Reflection and transmission of a wave incident on a slab with a
  time-periodic dielectric function $\epsilon(t)$}, Physical Review A
  \textbf{79}, 053821 (2009).

\bibitem{asadchy2022_time_dependent_scatterers}
V.~Asadchy, A.~Lamprianidis, G.~Ptitcyn, M.~Albooyeh, Rituraj, T.~Karamanos,
  R.~Alaee, S.~Tretyakov, C.~Rockstuhl, and S.~Fan, \textit{Parametric mie
  resonances and directional amplification in time-modulated scatterers},
  Physical Review Applied \textbf{18}, 054065 (2022).

\bibitem{rubinsztein2016roadmap_structured_light}
H.~Rubinsztein-Dunlop, A.~Forbes, M.~V. Berry, M.~R. Dennis, D.~L. Andrews,
  M.~Mansuripur, C.~Denz, C.~Alpmann, P.~Banzer, T.~Bauer \textit{et~al.},
  \textit{Roadmap on structured light}, Journal of Optics \textbf{19}, 013001
  (2016).

\bibitem{wheaton2015_modulation_EAR}
S.~Wheaton, R.~M. Gelfand, and R.~Gordon, \textit{Probing the {R}aman-active
  acoustic vibrations of nanoparticles with extraordinary spectral resolution},
  Nature Photonics \textbf{9}, 68 (2015).

\bibitem{chen2012optical_near_field_tip_plasmons_scanning_koppens}
J.~Chen, M.~Badioli, P.~Alonso-Gonz{\'a}lez, S.~Thongrattanasiri, F.~Huth,
  J.~Osmond, M.~Spasenovi{\'c}, A.~Centeno, A.~Pesquera, P.~Godignon, A.~Z.
  Elorza, N.~Camara, F.~J. García~de Abajo, R.~Hillenbrand, and F.~H.~L.
  Koppens, \textit{Optical nano-imaging of gate-tunable graphene plasmons},
  Nature \textbf{487}, 77 (2012).

\bibitem{fei2012_near_field_tip_plasmons_scanning_basov}
Z.~Fei, A.~Rodin, G.~O. Andreev, W.~Bao, A.~McLeod, M.~Wagner, L.~Zhang,
  Z.~Zhao, M.~Thiemens, G.~Dominguez, M.~M. Fogler, A.~H. Castro~Neto, C.~N.
  Lau, F.~Keilmann, and D.~N. Basov, \textit{Gate-tuning of graphene plasmons
  revealed by infrared nano-imaging}, Nature \textbf{487}, 82 (2012).

\bibitem{chaves2020_2d_semiconductors_bandgaps}
A.~Chaves, J.~G. Azadani, H.~Alsalman, D.~Da~Costa, R.~Frisenda, A.~Chaves,
  S.~H. Song, Y.~D. Kim, D.~He, J.~Zhou \textit{et~al.}, \textit{Bandgap
  engineering of two-dimensional semiconductor materials}, npj 2D Materials and
  Applications \textbf{4}, 1 (2020).

\bibitem{kim2015dirac_black_phosphorus}
J.~Kim, S.~S. Baik, S.~H. Ryu, Y.~Sohn, S.~Park, B.-G. Park, J.~Denlinger,
  Y.~Yi, H.~J. Choi, and K.~S. Kim, \textit{Observation of tunable band gap and
  anisotropic dirac semimetal state in black phosphorus}, Science \textbf{349},
  723 (2015).

\bibitem{chaves2017_excitonic_tmdcs}
A.~Chaves, R.~Ribeiro, T.~Frederico, and N.~Peres, \textit{Excitonic effects in
  the optical properties of 2d materials: an equation of motion approach}, 2D
  Materials \textbf{4}, 025086 (2017).

\bibitem{seetharam2015baths_controlled_floquet_population}
K.~I. Seetharam, C.-E. Bardyn, N.~H. Lindner, M.~S. Rudner, and G.~Refael,
  \textit{Controlled population of {F}loquet-{B}loch states via coupling to
  {B}ose and {F}ermi baths}, Physical Review X \textbf{5}, 041050 (2015).

\bibitem{Note1}
Note that $k_{F}$ is fixed by momentum conservation.

\bibitem{abrikosov1959}
A.~A. Abrikosov and I.~M. Khalatnikov, \textit{The theory of a fermi liquid
  (the properties of liquid 3he at low temperatures)}, Reports on Progress in
  Physics \textbf{22}, 329 (1959).

\bibitem{Eguiluz1976hydrodynamicPlasmons}
A.~Eguiluz and J.~Quinn, \textit{Hydrodynamic model for surface plasmons in
  metals and degenerate semiconductors}, Physical Review B \textbf{14}, 1347
  (1976).

\bibitem{Forster}
D.~Forster, \textit{Hydrodynamic Fluctuations, Broken Symmetry, and Correlation
  Functions}, CRC Press (2018). ISBN 978-0367091323.

\bibitem{lucas2015memory}
A.~Lucas and S.~Sachdev, \textit{Memory matrix theory of magnetotransport in
  strange metals}, Physical Review B \textbf{91}, 195122 (2015).

\bibitem{kiselev2021_superdiffusive_modes}
E.~I. Kiselev, \textit{Universal superdiffusive modes in charged two
  dimensional liquids}, Physical Review B \textbf{103}, 235116 (2021).

\bibitem{else2020discrete_time_cryst_review}
D.~V. Else, C.~Monroe, C.~Nayak, and N.~Y. Yao, \textit{Discrete time
  crystals}, Annual Review of Condensed Matter Physics \textbf{11}, 467 (2020).

\bibitem{else2016floquet_discrete_time_crystals}
D.~V. Else, B.~Bauer, and C.~Nayak, \textit{Floquet time crystals}, Physical
  review letters \textbf{117}, 090402 (2016).

\bibitem{yao2017discrete_time_crystal_original}
N.~Y. Yao, A.~C. Potter, I.-D. Potirniche, and A.~Vishwanath, \textit{Discrete
  time crystals: Rigidity, criticality, and realizations}, Physical review
  letters \textbf{118}, 030401 (2017).

\bibitem{zhang2017observation_discrete_time_crystal}
J.~Zhang, P.~W. Hess, A.~Kyprianidis, P.~Becker, A.~Lee, J.~Smith, G.~Pagano,
  I.-D. Potirniche, A.~C. Potter, A.~Vishwanath \textit{et~al.},
  \textit{Observation of a discrete time crystal}, Nature \textbf{543}, 217
  (2017).

\bibitem{kyprianidis2021observation_discrete_time_crystal}
A.~Kyprianidis, F.~Machado, W.~Morong, P.~Becker, K.~S. Collins, D.~V. Else,
  L.~Feng, P.~W. Hess, C.~Nayak, G.~Pagano \textit{et~al.}, \textit{Observation
  of a prethermal discrete time crystal}, Science \textbf{372}, 1192 (2021).

\bibitem{natsheh2021critical_properties_time_crystal}
M.~Natsheh, A.~Gambassi, and A.~Mitra, \textit{Critical properties of the
  prethermal floquet time crystal}, Physical Review B \textbf{103}, 224311
  (2021).

\bibitem{yao2020classical_discrete_time_crystal}
N.~Y. Yao, C.~Nayak, L.~Balents, and M.~P. Zaletel, \textit{Classical discrete
  time crystals}, Nature Physics \textbf{16}, 438 (2020).

\bibitem{ni2018_plasmon_quality_factor_graphene}
G.~Ni, d.~A. McLeod, Z.~Sun, L.~Wang, L.~Xiong, K.~Post, S.~Sunku, B.-Y. Jiang,
  J.~Hone, C.~R. Dean \textit{et~al.}, \textit{Fundamental limits to graphene
  plasmonics}, Nature \textbf{557}, 530 (2018).

\bibitem{ince1956o_ODE_book}
E.~L. Ince, \textit{Ordinary differential equations}, Courier Corporation
  (1956).

\bibitem{Note2}
The quantity $\epsilon _{\protect \mathrm {pl}}\left (q\right )$ is analogous
  to the quasi-energy of floquet driven electrons.

\bibitem{gerry2005introductory_quantum_optics}
C.~Gerry, P.~Knight, and P.~L. Knight, \textit{Introductory quantum optics},
  Cambridge university press (2005).

\bibitem{sun2022graphene_entangled_plasmon_pairs}
Z.~Sun, D.~Basov, and M.~Fogler, \textit{Graphene as a source of entangled
  plasmons}, Physical Review Research \textbf{4}, 023208 (2022).

\bibitem{burns2020dedalus}
K.~J. Burns, G.~M. Vasil, J.~S. Oishi, D.~Lecoanet, and B.~P. Brown,
  \textit{Dedalus: A flexible framework for numerical simulations with spectral
  methods}, Physical Review Research \textbf{2}, 023068 (2020).

\bibitem{principi2013intrinsic_graphene_plasmon_quality_factor}
A.~Principi, G.~Vignale, M.~Carrega, and M.~Polini, \textit{Intrinsic lifetime
  of dirac plasmons in graphene}, Physical Review B \textbf{88}, 195405 (2013).

\bibitem{rudner2020band_engineering}
M.~S. Rudner and N.~H. Lindner, \textit{Band structure engineering and
  non-equilibrium dynamics in floquet topological insulators}, Nature reviews
  physics \textbf{2}, 229 (2020).

\bibitem{pertsova2018excitonic_instability}
A.~Pertsova and A.~V. Balatsky, \textit{Excitonic instability in optically
  pumped three-dimensional {D}irac materials}, Physical Review B \textbf{97},
  075109 (2018).

\bibitem{sachdev1998_elliptic_coordinates}
S.~Sachdev, \textit{Nonzero-temperature transport near fractional quantum hall
  critical points}, Physical Review B \textbf{57}, 7157 (1998).

\bibitem{baker2013cooling_power_graphene_exp2}
A.~Baker, J.~Alexander-Webber, T.~Altebaeumer, S.~McMullan, T.~Janssen,
  A.~Tzalenchuk, S.~Lara-Avila, S.~Kubatkin, R.~Yakimova, C.-T. Lin
  \textit{et~al.}, \textit{Energy loss rates of hot {D}irac fermions in
  epitaxial, exfoliated, and cvd graphene}, Physical Review B \textbf{87},
  045414 (2013).

\bibitem{baker2012_cooling_power_graphene_exp1}
A.~Baker, J.~Alexander-Webber, T.~Altebaeumer, and R.~Nicholas, \textit{Energy
  relaxation for hot dirac fermions in graphene and breakdown of the quantum
  hall effect}, Physical review B \textbf{85}, 115403 (2012).

\bibitem{betz2012_cooling_power_graphene_exp_values}
A.~Betz, F.~Vialla, D.~Brunel, C.~Voisin, M.~Picher, A.~Cavanna, A.~Madouri,
  G.~F{\`e}ve, J.-M. Berroir, B.~Pla{\c{c}}ais \textit{et~al.}, \textit{Hot
  electron cooling by acoustic phonons in graphene}, Physical Review Letters
  \textbf{109}, 056805 (2012).

\bibitem{kaasbjerg2014_cooling_power_tmds}
K.~Kaasbjerg, K.~Bhargavi, and S.~Kubakaddi, \textit{Hot-electron cooling by
  acoustic and optical phonons in monolayers of mos 2 and other
  transition-metal dichalcogenides}, Physical Review B \textbf{90}, 165436
  (2014).

\bibitem{kubakaddi2009_cooling_power_graphene_theory}
S.~Kubakaddi, \textit{Interaction of massless {D}irac electrons with acoustic
  phonons in graphene at low temperatures}, Physical Review B \textbf{79},
  075417 (2009).

\end{thebibliography}
\end{document}